\RequirePackage[hyphens]{url}
\documentclass[12pt]{article}
\usepackage{booktabs}
\usepackage{natbib}
\usepackage{array}
\newcolumntype{L}{>{\centering\arraybackslash}m{3cm}}
\usepackage{braket}
\usepackage{microtype}
\usepackage[utf8]{inputenc}
\usepackage[bottom]{footmisc}
\usepackage{graphicx}
\usepackage{mwe}
\usepackage{amsmath}
\usepackage{amssymb}
\usepackage{tabularx}
\usepackage[capposition=top]{floatrow}
\usepackage{rotating}
\usepackage{dcolumn}
\usepackage{multirow}
\usepackage{nicefrac}
\usepackage{float}
\usepackage{placeins}
\usepackage{pdflscape}
\usepackage[singlespacing,nodisplayskipstretch]{setspace}
\usepackage{color}
\usepackage[margin=1in]{geometry}
\usepackage{varioref}
\usepackage{eurosym}
\usepackage[pdfpagemode=UseThumbs,
colorlinks=true,
urlcolor=black,
citecolor=black,
linkcolor=black]{hyperref}
\usepackage{rotating}
\usepackage{xcolor}
\usepackage{bookmark}
\usepackage[utf8]{inputenc}
\usepackage[english]{babel}
\usepackage{dirtytalk}
\usepackage[width=.75\textwidth]{caption}
\usepackage{subcaption}
\usepackage [autostyle, english = american]{csquotes}
\usepackage{lscape}
\MakeOuterQuote{"}
\usepackage{pdfpages}

\let\orgautoref\autoref
\providecommand{\Autoref}
{\def\equationautorefname{Eq.}%
	\def\figureautorefname{Figure}%
	\def\subfigureautorefname{Figure}%
	\def\sectionautorefname{Section}%
	\def\subsectionautorefname{Section}%
	\def\subsubsectionautorefname{Section}%
	\def\Itemautorefname{Item}%
	\def\tableautorefname{Table}%
	\orgautoref}

\renewcommand{\autoref}
{\def\equationautorefname{Equation}%
	\def\figureautorefname{Figure}%
	\def\subfigureautorefname{Figure}%
	\def\sectionautorefname{section}%
	\def\subsectionautorefname{section}%
	\def\subsubsectionautorefname{section}%
	\def\Itemautorefname{item}%
	\def\tableautorefname{Table}%
	\orgautoref}

\newcolumntype{d}[1]{D{.}{.}{#1}}
\newcolumntype{e}{D{.}{.}{-1}}
\newcolumntype{.}{D{.}{.}{3}}
\definecolor{navy}{rgb}{0,0,.5}

\graphicspath{{figs/}}

\begin{document}

\begin{spacing}{1.0}	

%Previous title: Bridging America's Divide on ...
	
\title{Reducing Polarization on \\Abortion, Guns and Immigration:\vspace{0.1 cm} \\An Experimental Study\vspace{0.1 cm}\vspace{0.2cm}\thanks{The study was funded by Cornell University (Belot's research funds). The authors thank Pilar Cardinale and Sarahi Osuna for excellent research support, and Nathan Maddox for very useful comments. We also thank the participants to the ESA 2022 Conference in Bologna for useful comments. This study obtained IRB approval from Cornell University, protocol no.: 2105010352. The experiment and the analyses were pregistered on the AEA RCT Registry, ID: AEARCTR-0008029.} 
\vspace{0.1cm}}

\author{Mich\`ele Belot\thanks{Cornell University \href{mailto:mb2693@cornell.edu }{\texttt{mb2693@cornell.edu}}.}
\and Guglielmo Briscese \thanks{University of Chicago, \href{mailto:gubri@uchicago.edu }{\texttt{gubri@uchicago.edu}}.}
\vspace{0.1cm}}
\maketitle

\begin{abstract}
  \noindent 
We study individuals' willingness to engage with others who hold opposite views on  polarizing policies. A representative sample of 2,507 Americans are given the opportunity to listen to recordings of fellow countrymen and women expressing their views on immigration, abortion laws and gun ownership laws. We find that most Americans (more than two-thirds) are willing to listen to a view opposite to theirs, and a fraction (ten percent) reports changing their views as a result. We also test whether emphasizing having common grounds with those who think differently helps bridging views. We identify principles the vast majority of people agree upon: (1) a set of fundamental human rights, and (2) a set of simple behavioral etiquette rules. A random subsample of people are made explicitly aware they share common views, either on human rights (one-third of the sample) or etiquette rules (another one-third of the sample), before they have the opportunity to listen to different views.  We find that the treatments induce people to adjust their views towards the center on abortion and immigration, relative to a control group, thus reducing polarization.

\end{abstract}
\bigskip



\end{spacing}

\begin{spacing}{1.5}

\newpage

\section{Introduction}

\noindent On June 24th 2022 the Supreme Court overturned the landmark ruling "Roe vs Wade" that recognized women's constitutional right to abortion, triggering significant reactions from both pro-choice and pro-life advocates. In the same week, the US Congress passed a gun control bill - described as the most significant firearms legislation in nearly 30 years - after two mass shootings in Buffalo, New York, and a primary school in Uvalde, Texas, that resulted in 31 deaths. The new measures were passed by 234 to 193 votes, a testimony to the polarized views on gun ownership laws.

%On June 24th 2022 the Supreme Court overturned the landmark ruling "Roe vs Wade" that recognized women's constitutional right to abortion. The decision triggered a massive reaction both from pro-choice advocates, voicing loudly their fury, and from pro-life supporters, celebrating the victory. In the same week, the US Congress passed a gun control bill - described as the most significant firearms legislation in nearly 30 years. The bill came after mass shootings in May 2022 at a supermarket in Buffalo, New York, and a primary school in Uvalde, Texas, that resulted in 31 deaths. The new measures were passed by 234 to 193 votes, with the debate on gun laws being fierce and alive. 

These significant developments are part of a seemingly long-term trend of increased division among Americans on important policies, reflecting a growing concern among political scientists about rising political polarization in the US and other Western societies \citep{bonomi2021identity, Mason2018}.\footnote{A global Ipsos study, carried out in 27 countries for the BBC in 2018 finds that three in four people on average across the 27 countries (76\%) think society in their country is divided\footnote{see https://www.ipsos.com/en-uk/bbc-global-survey-world-divided for full report}.} Recent studies point at polarization also becoming "affective", meaning ordinary Americans see other fellow citizens holding opposite political ideologies as hypocritical, selfish, and closed-minded \citep{Iyengar2019}. The concern is that American partisans are speaking `different languages', misunderstanding one another, and distrusting each other on a basic level \citep{Mason2018}. In response to these concerns, a number of initiatives have emerged to bridge the divide. In the US, programs such as "One Small Step" or "Braver Angels" bring together Americans of different political affiliations to exchange their views in workshops, debates or one-on-one conversations. 

The assumption that interactions and \textit{contact} could bridge differences is rooted in a long tradition of work in social and political psychology, such as  \citet{Allport1954} and other more recent  studies (see \citet{pettigrew2006meta}, \citet{pettigrew2005allport}, \citet{zhou2019extended} for meta-analyses), which mostly support the hypothesis that `contact' can bridge opposing views and attenuate prejudice. Other studies have also shown that brief interactions can have lasting effects on voting preferences and religion-based prejudice in conflict-prone areas \citep{pons2018will, scacco2018can}. This wealth of evidence raises the question:  Why have we not been successful in bridging the divide? We see two major challenges. First, while engaging with someone who thinks differently may help bridging views, social interactions are often endogenous: Most of us choose with whom to interact, and even if there are opportunities to talk and listen to individuals from other social groups, people tend to favor interactions with others who are more similar to them \citep{McPherson2001}. Further, simply articulating one's opinion on a topic might reinforce one's beliefs even when this is incorrect or arbitrary, thus limiting the potential presented by an exchange of opinions \citep{schwardmann2022self}. A second reason is that there might be specific conditions for contact to be beneficial. According to Allport's theory: \textit{"The effect [of contact] is greatly enhanced if this contact is sanctioned by institutional supports (i.e., by law, custom or local atmosphere), and provided it is of a sort that leads to the perception of common interests and common humanity between members of the two groups."} (see discussions in \citet{dixon2005beyond} and \citet{mckeown2017contact}). That is, a key condition for bridging views may be to first establish common grounds.

This paper aims to address these two challenges. First, in our setting, people will have the \textit{opportunity to engage} with others who think differently, rather than being "forced" to interact. We conduct an online study where participants are offered the opportunity to listen to short recordings of others expressing their views on policies that are known for being polarizing - namely, laws concerning abortion, gun ownership, and immigration. Participants are aware that these individuals hold different views than theirs and can choose to listen to them or not. Subsequently, participants are asked how the recordings affected their views. The voluntary character of the interaction may well be key in triggering a real willingness to listen. There is indeed a very well-established literature in psychology around the concept of “psychological reactance” \citep{brehm1966}, which states that if individuals feel their freedom is reduced or threatened, they may be motivationally aroused to regain them. In other words, people may be more open to changing their views if they are choosing whether to engage or not. Importantly, this setting allows us to study directly the willingness to engage with others who think differently. 

Our second contribution is to test an intervention targeting the key condition outlined in Allport's work: creating a sense of \textit{common humanity}. Previous studies attempted to create common goals or tasks by crafting situations that transform participants' perceptions of the memberships from “us” and “them” to a more inclusive “we” \citep{sherif1988robbers}, \citep{Gaertner1996}. These studies show that perceptions of a superordinate identity reduces outgroup bias. We propose instead an intervention that attempts to emphasize the "common humanity" more directly. In our view, there is no more direct translation of humanity than basic human rights. We first elicit views on a selected set of human rights statements as written in the United Nations Universal Declaration of Human Rights. As expected, the vast majority of people, independently of their race, gender or political affiliations, agrees with these basic human rights. We then introduce a treatment where we inform participants that they share the same views on human rights as those who expressed a view different than theirs on abortion laws, gun laws or immigration. This offers, in our view, a unique opportunity to foster a sense of common humanity.     

Of course, it is possible that the mere emphasis of common attributes or interests, no matter how futile they may be, could trigger a higher willingness to listen to others. The "minimal group paradigm" shows that even attributing irrelevant group identity tokens to individuals create a sense of belonging and in-group and out-group dynamics (\citet{Tajfel1971}, with a recent review of experimental findings in \citet{balliet2014ingroup}). To have a broader sense of how the impact of raising awareness about commonality may affect willingness to engage, we introduce a treatment where we inform participants that they share views on other, more futile, attitudes, specifically basic behavioral etiquette rules. 

We conduct the experiment with a representative sample of 2,507 US citizens, and collected audio recordings from a separate sample. We find a very high baseline level of Americans' willingness to listen: two-thirds of respondents are willing to listen to the recording of someone holding views opposite to their own, with the treatments not significantly shifting this tendency. Similarly, in the aggregate, views don't change after having had the opportunity to listen to someone holding opposing beliefs, but at the individual level 10\% of people do change their views. Emphasizing common grounds contributes to reducing polarization on the topic of abortion laws and to some extent immigration, with more extreme opinions shifting towards the center. These treatments are however ineffective in reducing polarization on gun laws, possibly because of lower baseline levels of polarization on this policy topic.    

The rest of the paper is organized as follows. Section \ref{sec:design} describes the experimental design and Section \ref{sec:Results} presents the results. Section \ref{sec:conclusion} concludes.

\section{Experimental Design}\label{sec:design}

\subsection{Providing an Opportunity to Engage}

A key feature of our setting is to allow people to choose to engage with others who think differently. We chose to operationalize this in the following way: Participants have the opportunity to listen to a short recording of someone expressing a view opposite to their own on a policy topic known to be divisive. Listening is obviously a specific form of engaging, but in our view, it is a crucial one. It is a prime mode of interaction in the current world, with people choosing what material to read or listen to off or on-line. We chose to use recordings because previous studies have shown that listening to another person's voice might be a more effective medium of communication compared to for example reading \citep{schroeder2017humanizing}. Further, we want to know whether listening to someone expressing their (opposite) opinion on a policy leads to a change in views. 

Note that we chose to give a choice to engage with someone with a different view or not, and did not offer the alternative of engaging with someone with a similar view instead, or any other alternative. The reason for not giving an alternative is because it is presumably easy for participants to access people who share their views in real life, while it is presumably much harder to access people who don't. 

The next section describes how we chose the \textit{common grounds} used as treatments and the policies that were more likely to be polarizing. Section \ref{sec:audios} describes the procedure for collecting recordings of Americans expressing their views about the policies (immigration, gun laws, and abortion). Finally, Section \ref{sec:main_experiment} describes the set up of the main experiment and the methodology for measuring the outcomes of interest. 

\subsection{Common Values}\label{sec:commonality}
\par Our treatments involve creating a sense of common interests and humanity. To do that, we first needed to identify values and principles people most agree on. We began by shortlisting 14 statements about human rights from the United Nations Human Rights Declaration (UNHRD) and 10 behavioral etiquette rules. The UNHRD, first proclaimed by the United Nations General Assembly in 1948, sets out fundamental human rights to be universally protected and is recognized to this date as the inspiration for human rights treaties around the world. The UNHRD compromises a total of 30 articles, from which we selected the shortest 14 to meet standard survey completion time constraints.\footnote{For a copy of the full declaration, see: \url{https://www.un.org/en/about-us/universal-declaration-of-human-rights}} The behavioral etiquette rules were selected from a private website providing information and education for refugees, asylum seekers, immigrants and welcoming communities living in the United States.\footnote{See: \url{https://usahello.org/life-in-usa/culture/be-polite/}} 

We then recruited a representative sample of 325 Americans and asked them to rank, on a slider scale from 0 to 10, how much they agreed or disagreed with each of the 24 statements.\footnote{Respondents took on average 4 minutes to complete the survey, and they were compensated for their time as per Qualtrics panel payment agreement. All statements were shown randomly to avoid fatigue or anchoring effects.} Note that we did not mention that these statements were derived from the United Nations Declaration of Human Rights. We then selected the statements that had the highest level of agreement as defined by the highest share of respondents who gave a score above 7 or below 3, as per preregistered trial protocol. In addition, we performed a series of tests that confirmed there was indeed no significant difference in the share of agreement on these statements between (self-declared) Democrats and Republicans or across socio-economic and demographic variables that were collected at the end of the survey (see Online Appendix for survey instructions and balance tests). 

The shortlist of human rights selected for the main experiment includes: \textit{(1) No one shall be held in slavery or servitude; slavery and slave trade shall be prohibited in all their forms; (2) No one shall be arbitrarily deprived of his property; (3) Everyone has the right to freedom of peaceful assembly and association; (4) Everyone has the right to life, liberty and security of person; (5) Everyone has the right to freely to participate in the cultural life of the community, to enjoy the arts and to share in scientific advancement and its benefits.} 

The shortlist of etiquette rules selected for the main experiment includes: \textit{(1) "Wait your turn in a waiting line"; (2) "Say `please' when you ask for something"; (3) "Say `thank you' to acknowledge service, kindness, or the receipt of something"; (4) "Arrive on time. Not being late for classes, intimate gatherings, or appointments."; (5) "Refrain from talking loudly in quiet settings (or on your cell phone in movies, plays, or other quiet or focused communal settings)".}  

\subsection{Polarizing Policies}\label{sec:policies}
 We selected three policies that have been shown in several polls to be polarizing among Americans, and that are often debated by politicians and media outlets: abortion, immigration and gun ownership laws.\footnote{In a first phase, we also included free trade as a fourth topic, as pollsters and commentators argued it became a polarizing topic in recent American politics. The question read: \textit{"Free trade agreements between the US and other countries have generally been [A bad thing-neither good nor bad-a good thing scale]"}. However, we were unable to reach a minimum sample of audio files of respondents who were against free trade, and therefore decided to drop it from the main survey. We updated the pre-analysis plan in the trial registry prior to the launch of the main survey.} For consistency, we looked at the level of polarization on these policies based on the most recent Ipsos polls available at the time of the design of the experiment. A 2019 IPSOS poll found that while most Americans believed abortion should be legal in most or all cases, among Republicans only 35\% believed so, compared to 75\% of Democrats.\footnote{Source:\url{https://www.ipsos.com/sites/default/files/ct/news/documents/2019-06/ipsos_usat_abortion_topline_053119.pdf} last accessed May 2022.} A 2021 survey from the same company found the immigration was an equally divisive topic.\footnote{Source:\url{https://www.ipsos.com/en-us/news-polls/immigration-americans-favor-both-restrictions-and-reforms} last accessed May 2022.} Even before the COVID-19 pandemic, polls showed that 62\% of Republicans believed the number of immigrants allowed in the country should be decreased, compared to about 25\% of Democrats\footnote{Source:\url{https://www.ipsos.com/en-us/news-polls/americans-views-immigration-policy}}. On gun laws, a 2017 IPSOS poll found that 65\% of Democrats believed that they should be a lot stricter compared to a lower 29\% of Republicans.\footnote{Source:\url{https://www.ipsos.com/en-us/news-polls/npr-gun-control-2017-10} last accessed May 2022}

\subsection{Recordings of Views on Polarizing Policies}\label{sec:audios}
\par We collected audio recordings from Americans expressing their views about the selected policies with a second on-line survey, using the Qualtrics Survey Panel. Respondents were first asked to state their level of agreement with each of the shortlisted set of human rights and etiquette rules from the first study phase. Then, they were asked to indicate their views on the abortion, immigration, and gun laws. The policy questions were shown on a slider scale 0 to 10, with text prompts on each extreme of the sliding bar, and the survey was then programmed to identify the policy that the respondent felt most strongly about as the one they assigned the highest or lowest score of the three (and randomly to break ties).\footnote{The policy slider questions asked whether: Abortion should be \textit{[illegal-legal scale]} in most cases; Current gun laws in the United States are \textit{[too strict-about right-too lenient scale]}; Legal immigrants in the United States today \textit{[Burden the country by taking jobs, healthcare, and housing-strengthen the country through hard work and talent scale]}} Respondents were then shown a screen reminding them of their views on the selected policy, and were asked to "\textit{share an audio file of yourself expressing, in about 60 seconds or less, your views about the topic. How would you explain your point of view about this topic to someone else?}". Thus, each respondent was asked to record their views only about one policy. To help them with the recording and uploading of the audio files, we also showed respondents a page with user-friendly instructions depending on the device they were using. After uploading the audio file, respondents were asked a short set of questions about their socio-economic and demographic characteristics, as well as their political views (see Online Appendix for a copy of the survey). 

Respondents who completed the survey and correctly submitted an audio file received a larger compensation for their time compared to standard surveys.\footnote{Compensation was based on Qualtrics panel payment agreement.} This second survey was conducted over the Summer and early Fall of 2021. We received around 100 audio files that we screened to shortlist the ones that were at least 30 seconds long, were of sufficiently good sound quality that could be listened from any device, and did not contain profanity or swearing. No other exclusion criteria were applied. This selection processes gave us 47 audio files: 10 (7) against (pro) abortion, 4 (9) against (pro) immigration, and 9 (8) against (pro) stricter gun laws. 

\subsection{Main Experiment}\label{sec:main_experiment}

We launched the main survey through Qualtrics on April 4 and closed it on April 8, 2022, after reaching the pre-specified sample size of 2,507 Americans. Respondents who took the survey first saw a participant information statement (PIS) and consent form, which instructed them that they would be asked to answer a number of questions and be given the opportunity to listen to audio recordings (without specifying the substance of these recordings). Participants received a flat participation fee, regardless of how much time they spent on the survey. 

\begin{figure}[H]
   \centering
    \caption{Audio listening instructions text}
    \label{fig:instructions_audio}
   \includegraphics[height=8.5cm,angle=0]{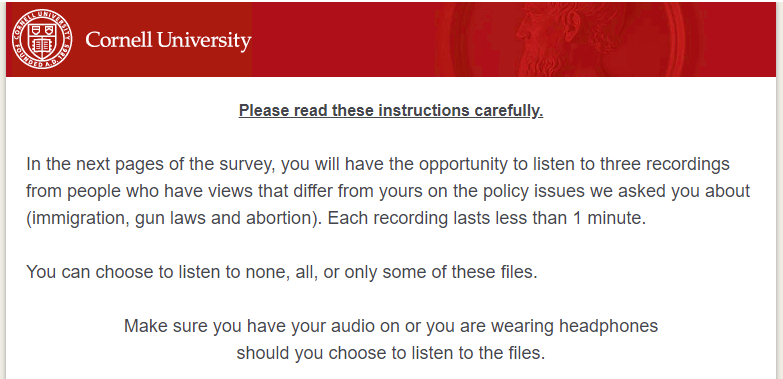}
  \end{figure}

The first survey questions asked respondents to rank, on a 0-to-10 slider scale, their agreement with the human rights and behavioral etiquette statements selected from the previous study phase, as well as their views on the three policies (immigration, abortion and gun laws). The subsequent part of the survey informed participants that they would have the opportunity to listen to recordings from people who have views that differed from theirs on the three policies. Participants were explicitly reminded that they could choose to listen to none, all, or only some of these files. The message also reminded respondents to make sure their volume or headphones were on. The purpose of this additional instruction page was to make sure that engagement with the audio file was not affected by any technical issue (e.g., malfunctioning sound) and that respondents understood that their engagement with the audio files was completely voluntary (see \Autoref{fig:instructions_audio} for a screenshot of the instructions message). 

Respondents were then randomly allocated to control or one of the treatment groups. The control group only saw a blank page asking them to click a button to continue to the next page. The \textit{"Etiquette Rules"} treatment group (hereafter, T1 group) saw a page reminding them of the etiquette questions they were just asked, and were told: "\textit{We asked these exact same questions to the people who recorded the audio files you will be shown in the next pages and each of them had a rating of agreement above 9 out of 10.}". We then showed them again the full list of etiquette rules as an additional reminder. Respondents who were randomly allocated to the \textit{Human Rights} treatment group (hereafter, T2 group) saw an almost identical page reminding them of the human rights questions they just replied to, and informing them that the people who recorded the audio files they were about to be shown had a rate of agreement above 9 out of 10 on those statements (see Online Appendix for details of the treatment messages). On average, respondents in the control group spent 8 seconds reading this instruction page, compared to about 24 seconds in T1 group and 19 seconds in the T2 group.\footnote{As explained in greater detail in the Results section, the difference between the two treatment groups was not significant; t=1.10, Satterthwaite's df=911, \textit{p}=0.267}

After reading the instructions, respondents landed on a page with the first recording. They had to click on the Play Button to listen to it. They were automatically matched with a recording of an American who expressed views opposed to theirs (as measured by their response to the same policy questions). For example, if someone stated to be against stricter gun laws at the beginning of the survey, we would show them an audio file of someone explaining why they thought gun laws in the United States were too lenient. As we had multiple audio files for each policy view, the recording was randomly selected from the relevant group to (e.g., recordings in favor of stricter gun laws). Further, we also randomized the order in which respondents saw the policies - that is, whether they first saw an audio file about immigration, gun laws, or abortion, since the audio files were shown one per page, sequentially. 

While we used a hidden timer to track how many seconds respondents spent on each page with an audio file, we also wanted to collect self-reported levels of engagement. Following the three pages displaying each of the recordings, respondents were asked to report, for each policy, (a) whether they listened to all, some, or none of the audio file, and (b) whether listening to the recordings changed their views about that policy (conditional on listening to it). If the respondent stated their views changed on a policy, they were asked again the same policy view question as at the beginning of the survey; if they said their views didn't change, they were asked to explain why.\footnote{The question asked: "Please indicate why your views on (policy) have not changed: (i) I already knew and considered the arguments presented, (ii) I didn't find the arguments convincing, (iii) other (free text)"} The last survey module collected respondents' socio-economic and demographic characteristics, as well as their political ideology. We purposely left these questions at the end to avoid priming respondents about their political identity. 

\section{Descriptive statistics}\label{sec:Results}

\subsection{Basic demographics}

We first present descriptive statistics of the sample participating in the main experimental study. The 2,507 Americans recruited through the Qualtrics Survey Panel were randomly allocated across control and treatment groups. The randomization ensured an almost perfect balance across all key demographic and socio-economic characteristics, in particular for political party affiliation which is perhaps the most important correlate of views on the policies we selected. Table \ref{t:balance} in Appendix \ref{sec:Appendix_tables} reports the averages of the main socio-economic and demographic characteristics, and the relative balance checks across groups. 

\subsection{Baseline levels of agreement}

\begin{figure}[H]
     \centering
     \begin{subfigure}[b]{0.45\textwidth}
         \centering
         \includegraphics[width=\textwidth]{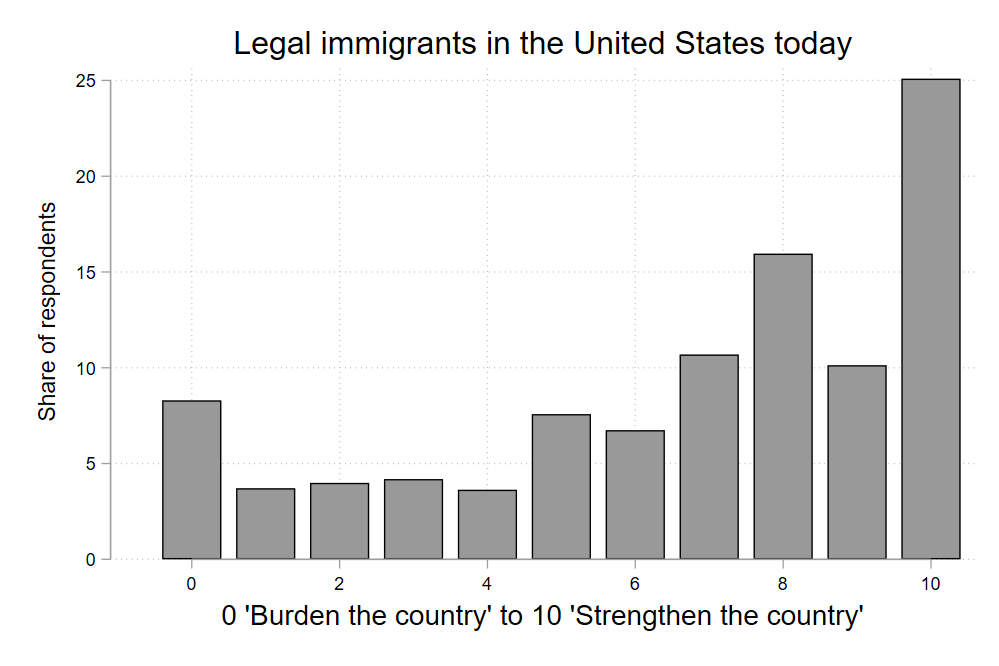}
         \label{fig:immi}
     \end{subfigure}
     \hfill
     \begin{subfigure}[b]{0.45\textwidth}
         \centering
         \includegraphics[width=\textwidth]{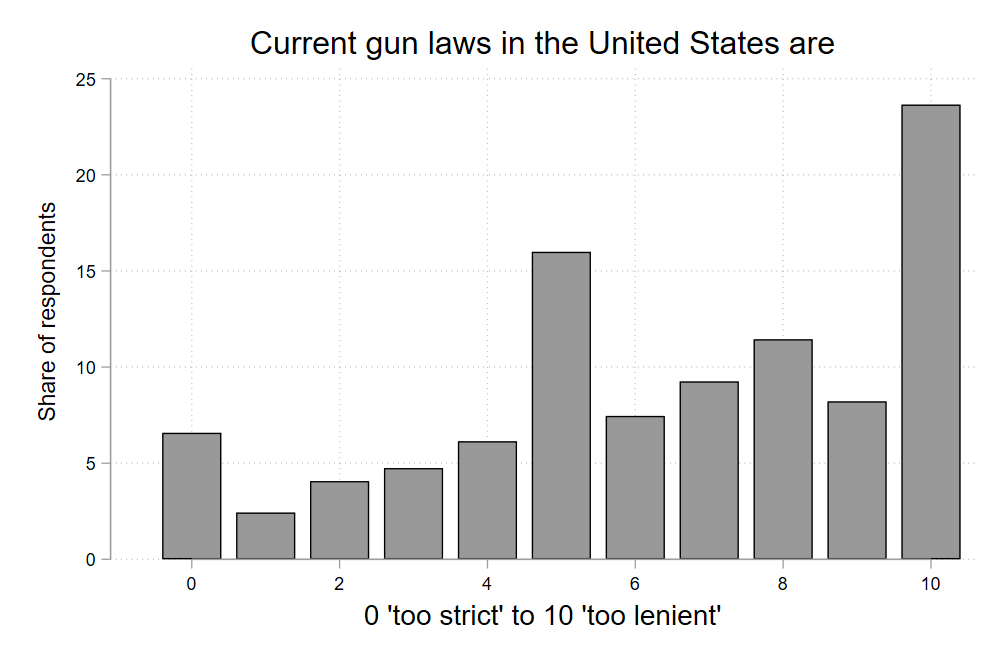}
         \label{fig:guns}
     \end{subfigure}
     \hfill
     \begin{subfigure}[b]{0.45\textwidth}
         \centering
         \includegraphics[width=\textwidth]{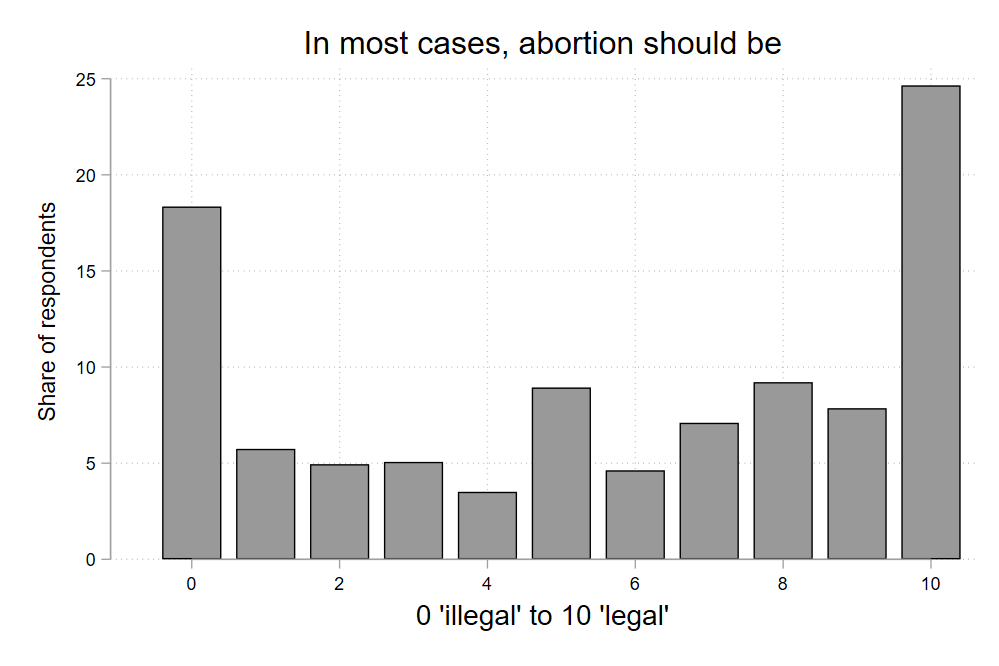}
         \label{fig:abortion}
     \end{subfigure}
        \caption{Baseline policy views}
        \label{fig:policy_baseline}
  \vspace{-10pt}
   \floatfoot{\textbf{\textit{Notes.}} The figure plots the share of respondents who gave a score between 0 and 10 to each of the three policy statements. The title and the x-axis label show the exact wording shown to respondents on the slider questions.}
\end{figure}

Figure \ref{fig:policy_baseline} shows the baseline distribution of attitudes towards the three selected policies. As expected, views on these topics are spread. On immigration, around 23\% of the sample gave a score between 0 and 4 out of 10, thus believing that legal immigrants in the United States are more of a burden than a strength; conversely, 25\% of respondents gave a score of exactly 10, strongly agreeing that legal immigrants strengthen the country. Views on abortion laws appear to be the most polarized, with more than 40\% of the sample answering at the very extremes (0 or 10), with an almost equal split of 18\% giving a score of 0, believing in most cases abortion should be illegal, and about 25\% giving a score of 10 (i.e., in most cases abortion should be legal). Gun laws appears to be the relatively least polarizing policy of the three, with about 16\% of respondents giving a score of 5 out of 10, although about 23\% giving a score between 0 and 4 (`too strict'), and the remaining 60\% giving a score between 6 and 10 (`too lenient').

\begin{figure}[H]
     \centering
     \begin{subfigure}[b]{0.45\textwidth}
         \centering
         \includegraphics[width=\textwidth]{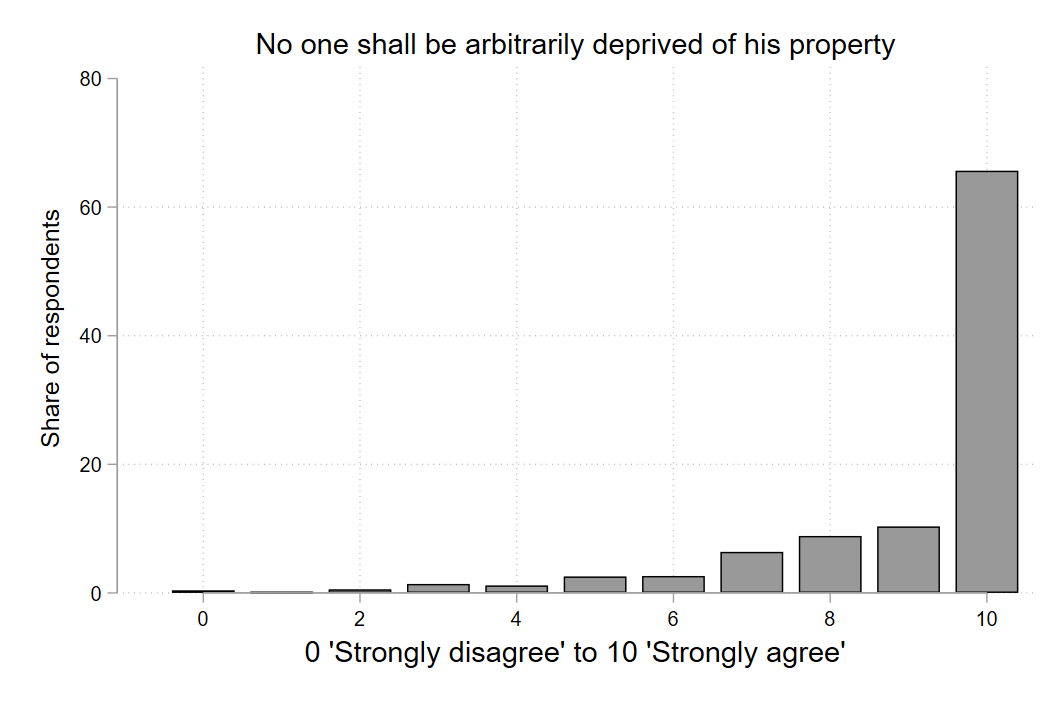}
         \label{fig:property}
     \end{subfigure}
     \hfill
     \begin{subfigure}[b]{0.45\textwidth}
         \centering
         \includegraphics[width=\textwidth]{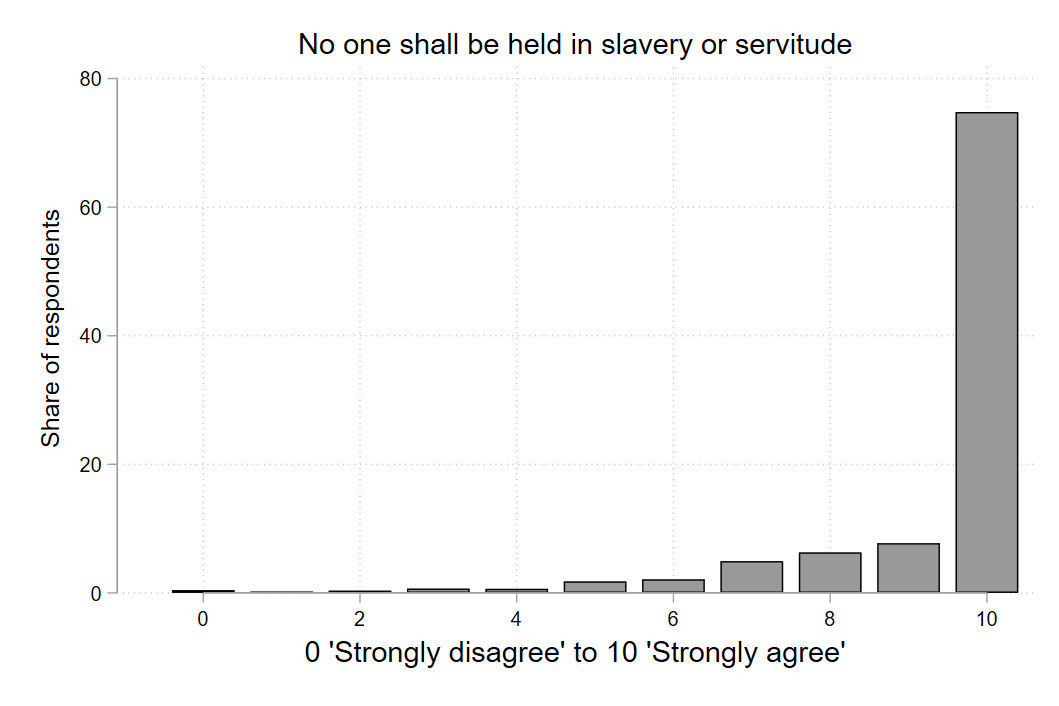}
         \label{fig:slavery}
     \end{subfigure}
     \hfill
     \begin{subfigure}[b]{0.45\textwidth}
         \centering
         \includegraphics[width=\textwidth]{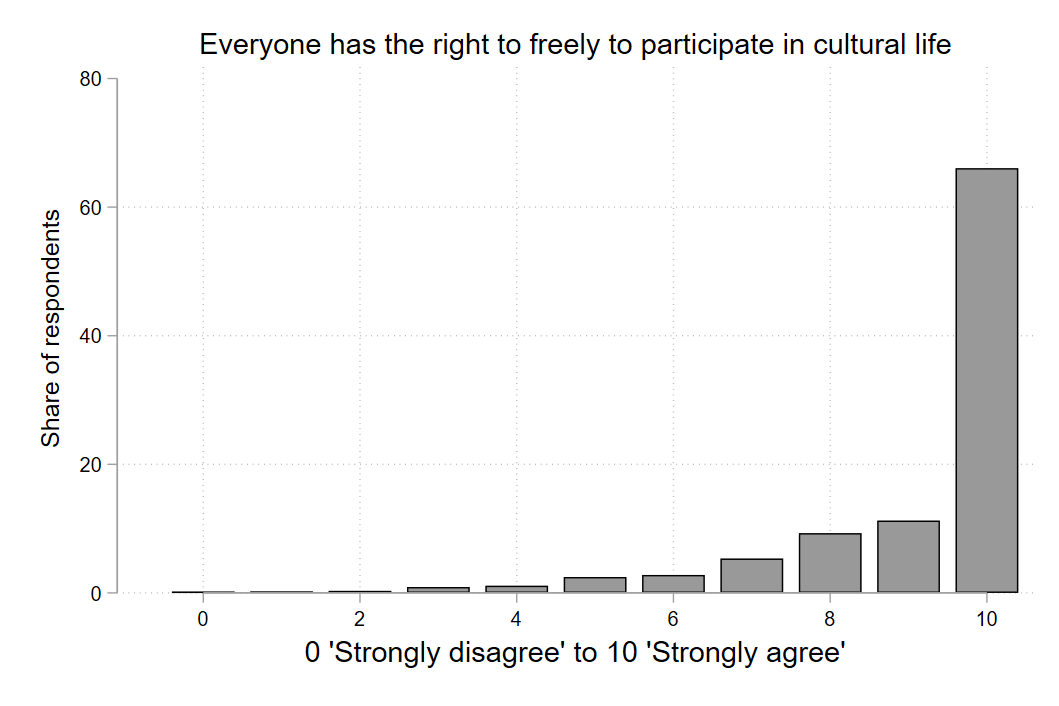}
         \label{fig:culture}
     \end{subfigure}
     \hfill
     \begin{subfigure}[b]{0.45\textwidth}
         \centering
         \includegraphics[width=\textwidth]{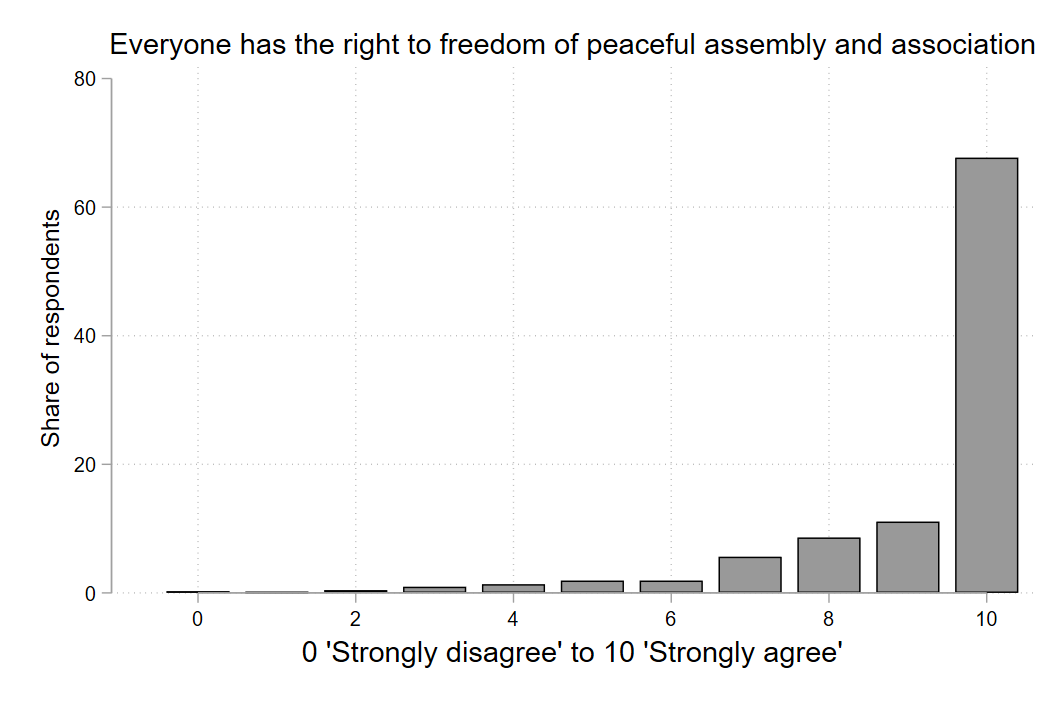}
         \label{fig:assembly}
     \end{subfigure}
    \hfill
     \begin{subfigure}[b]{0.45\textwidth}
         \centering
         \includegraphics[width=\textwidth]{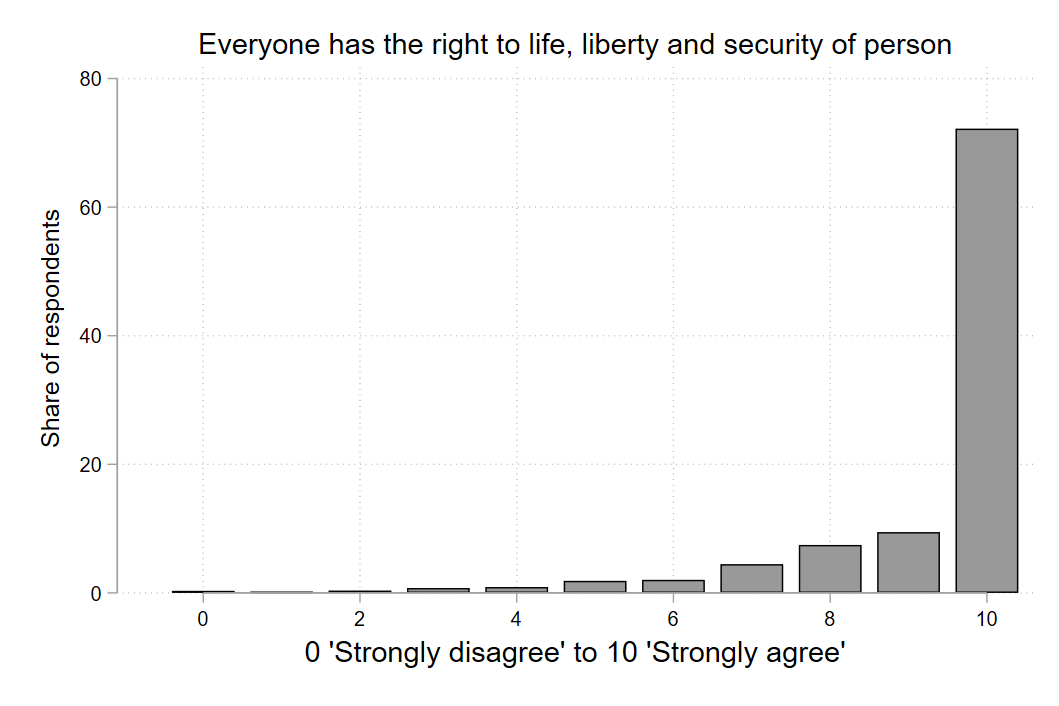}
         \label{fig:lifeliberty}
    \end{subfigure}
        \caption{Attitudes on human rights values}
        \label{fig:values}
  \vspace{-10pt}
   \floatfoot{\textbf{\textit{Notes.}} The figure plots the share of respondents who gave a score between 0 (strongly disagree) to 10 (strongly agree) to each of the five human rights values statements.}
\end{figure}

\begin{figure}[H]
     \centering
     \begin{subfigure}[b]{0.45\textwidth}
         \centering
         \includegraphics[width=\textwidth]{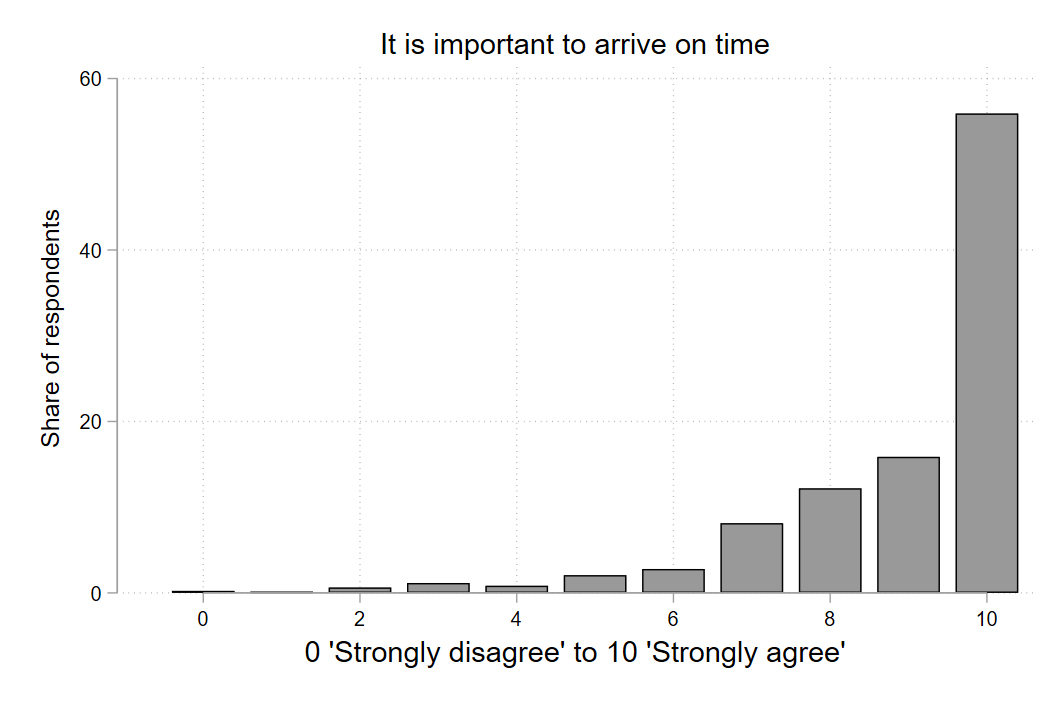}
         \label{fig:etiq_ontime}
     \end{subfigure}
     \hfill
     \begin{subfigure}[b]{0.45\textwidth}
         \centering
         \includegraphics[width=\textwidth]{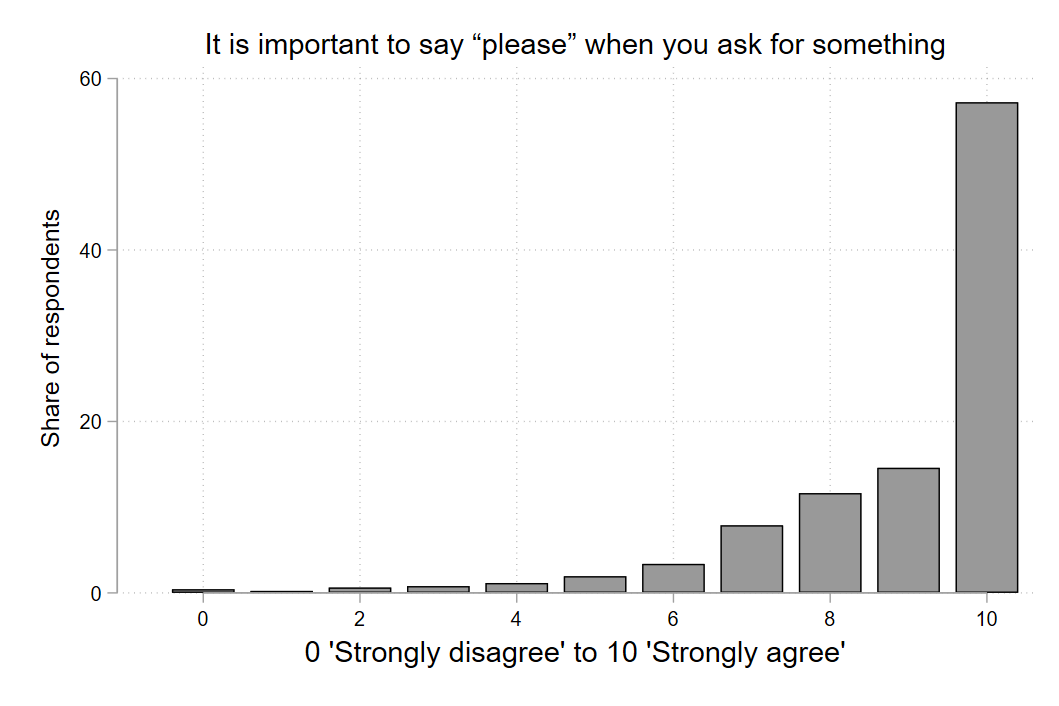}
         \label{fig:etiq_sayplease}
     \end{subfigure}
     \hfill
     \begin{subfigure}[b]{0.45\textwidth}
         \centering
         \includegraphics[width=\textwidth]{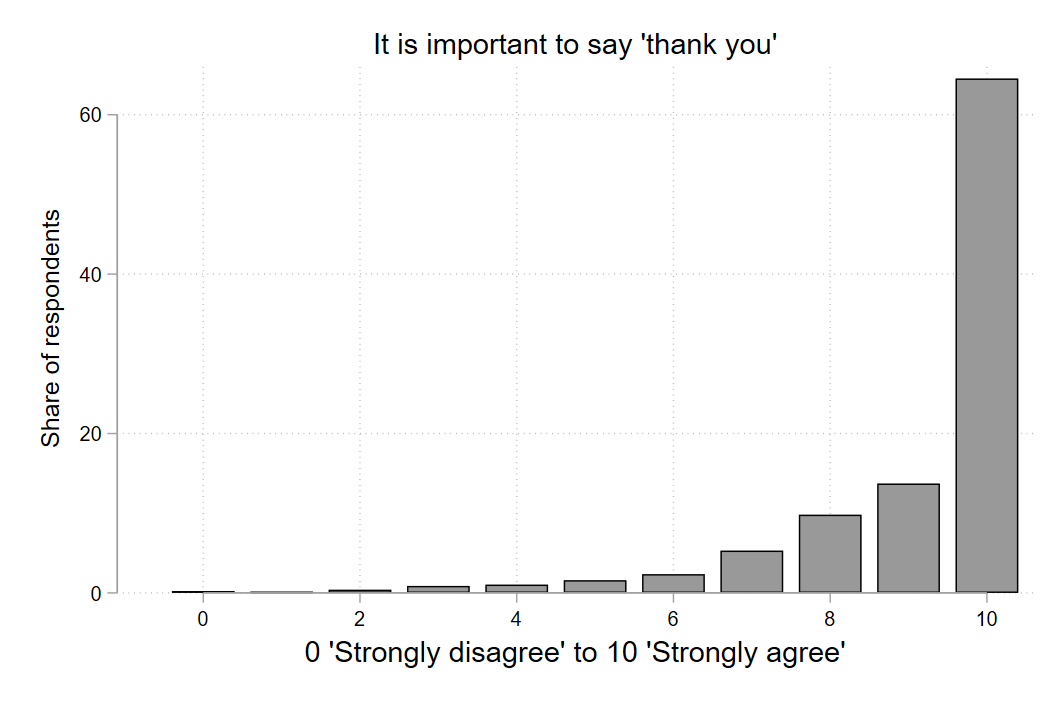}
         \label{fig:etiq_saythanks}
     \end{subfigure}
     \hfill
     \begin{subfigure}[b]{0.45\textwidth}
         \centering
         \includegraphics[width=\textwidth]{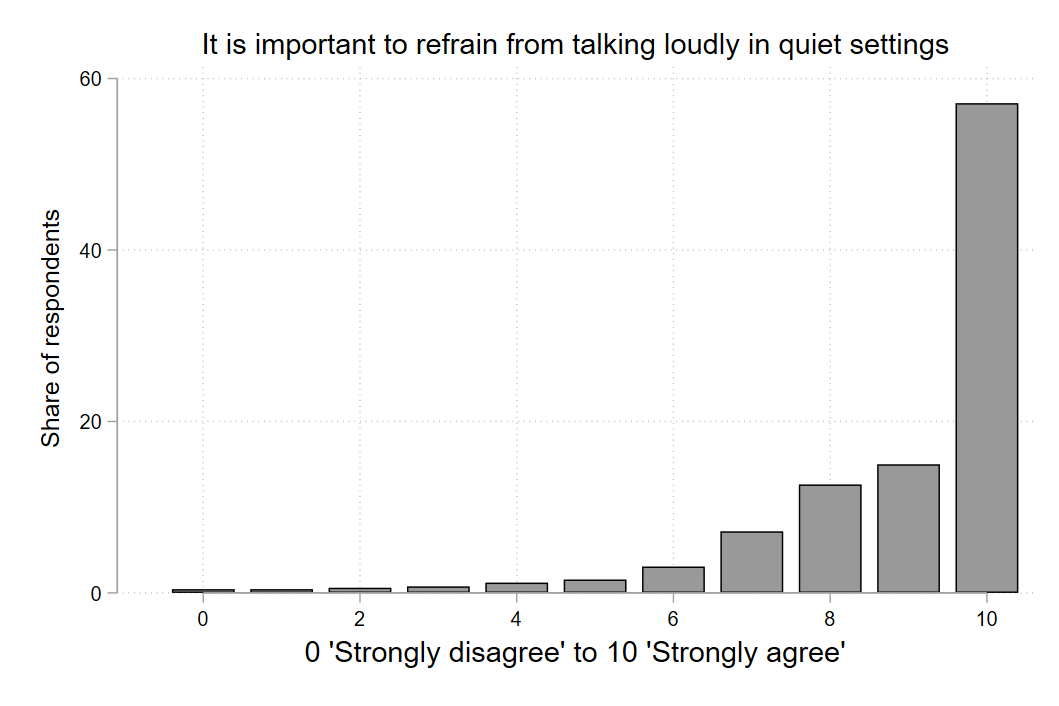}
         \label{fig:etiq_talkloud}
     \end{subfigure}
    \hfill
     \begin{subfigure}[b]{0.45\textwidth}
         \centering
         \includegraphics[width=\textwidth]{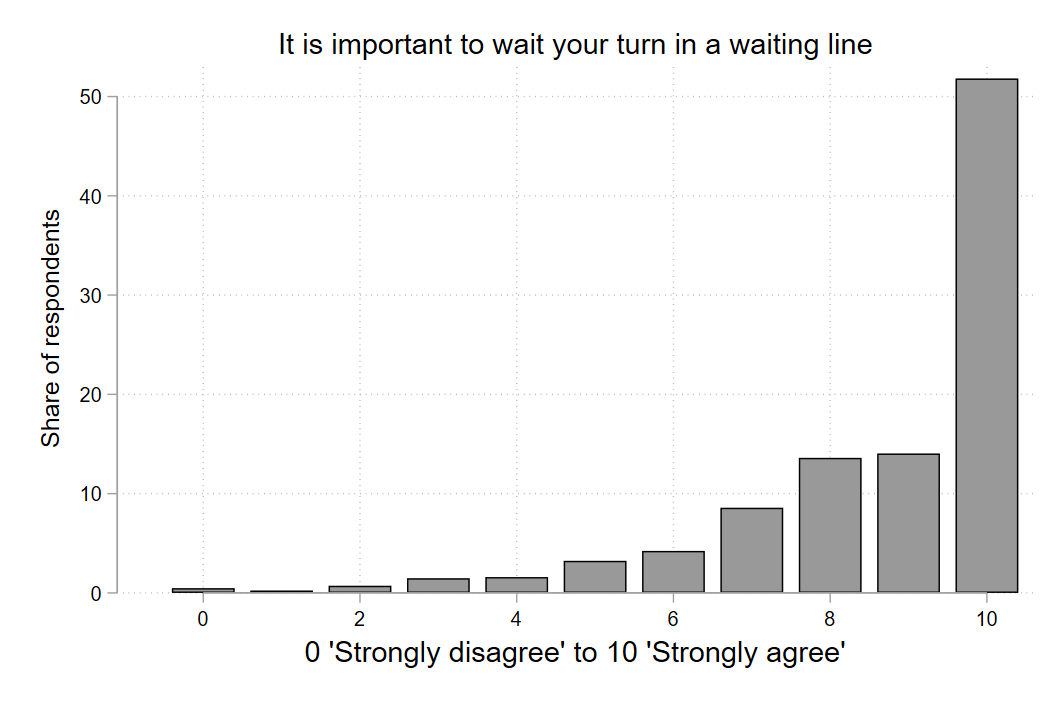}
         \label{fig:etiq_waitline}
    \end{subfigure}
        \caption{Attitudes on behavioral etiquette}
        \label{fig:etiquettes}
  \vspace{-10pt}
   \floatfoot{\textbf{\textit{Notes.}} The figure plots the share of respondents who gave a score between 0 (strongly disagree) to 10 (strongly agree) to each of the five behavioral etiquette statements.}
\end{figure}

In contrast, views on human rights and etiquette rules are, as expected, much more aligned. For each of the questions corresponding to either category, more than 80\% of the sample reports a rate of agreement of 8 or more on a scale from 0 to 10 (see Figures \ref{fig:values} and \ref{fig:etiquettes}).

\section{Results}\label{sec:Results_WTE}
\subsection{Willingness to engage}\label{sec:Results_WTE}
A key feature of our design is that participants are given an opportunity to listen to others, knowing they have a different view, but are not forced to do so (see \citet{minson2022receptiveness} for a review of studies using different methodologies to measure receptiveness to others). Our first outcome of interest is their willingness to engage (WTE). We defined two variables in our registered pre-analysis plan to capture this outcome. The first outcome is the number of recordings listened to (0, 1, 2 or 3). A recording is considered as `listened to' if the participant spent at least 5 seconds on the page where the recording appeared and indicated having listened to "some of it" or “all of it” following the recording. The second outcome is the percentage of recordings listened to, measured as the time spent on the pages with the audio recordings divided by the total time length of the recordings. 

\newpage

\textbf{Result 1. Levels of engagement are high.}\\
Our first result is that both measures indicate very high levels of engagement. The first measure indicates that 69\% of respondents listened to all three files, and about 18\% to two files. The second measure indicates that 56\% of the sample listened to the recordings for an amount of time that was equal to or greater than their total length (mean 1.03, s.d. 1.006). (see Figures \ref{fig:tot_perc_engage} and \ref{fig:self_reported_eng} in Appendix \ref{sec:Appendix_tables} for the distribution graphs). 

One concern is that participants had "nothing else to do" and therefore listened to the recordings for that reason. We made sure the instructions clearly stated that they had the choice to listen or not. Obviously, they could finish the survey faster if they would not spend time listening to recordings. So while it is true that we cannot be sure that participants \textit{really} listened to the actual substance of the recording, it is notable to see such level of engagement when it was clearly communicated to them that it was voluntary and there were no incentives to do so. 

To have a better sense of the degree to which survey participants usually engage with content, we ran an additional survey on a different representative sample of Americans, partnering with the same survey firm (Qualtrics). The goal here is to obtain a benchmark of level of engagement with audio recordings of a different kind. We recruited 110 Americans to take part in a survey on the 14th and 15th of December 2022. All survey questions and the instructions remained identical to the main survey, except for the questions about the three selected policies. There was no between-subjects randomization, i.e. the survey was identical for everyone. Participants also had the opportunity to listen to up to three audio files of approximately 1 minute length, but instead of being on one of the polarizing policy topics selected for the main experiment, these audio files were on topics of general interest extracted from the Scientific American's online podcast `60-Second Science'.\footnote{For more information: \url{https://www.scientificamerican.com/podcast/60-second-science/}} The contents of these recordings relate to scientific and academic research, including astronomy, zoology, and history. As in the main survey, each audio file page was accompanied by a brief one-sentence explanation of the topic discussed in the audio and a hidden timer was embedded in each page (see survey screenshots in \ref{sec:Appendix_instructions} for an example). We find almost identical levels of the willingness to engage in our main survey and in this additional comparison survey. That is, in general the WTE is high, and the level of engagement we see in our main experiment does not appear particularly high or low (see \Autoref{t:wte_baseline} for a summary comparison table). 

\textbf{Result 2. Emphasizing common ground does not further increase willingness to engage.}\\
Our second result evaluates the effects of the two treatments on the WTE outcomes. For the first outcome we report results from a negative binomial regression, and for the second outcome we report the results of a Tobit regression since the outcome variable is truncated at 100\%, and 56\% of our sample has a value of 100\%.\footnote{In the pre-analysis plan we announced we would run a linear regression, but did not anticipate that more than half of our sample would be truncated at the top. Given the data structure, a Tobit model is more appropriate.} \Autoref{t:wte} presents the estimated treatment effects. We find precisely estimated null effects, that is, we do not find any indication that emphasizing common grounds increases people's willingness to listen. Knowing that the other person shares the same views on behavioral etiquette rules (T1/etiq) increases the willingness to listen compared to the control group by a small and non-significant 0.007 points on our first outcome measure, and increases also in a non-significant manner the percentage of audio recordings listened to by around one percent compared to the control group. Knowing that the other person shares the same views on human rights (T2/values) decreases in a non-significant way respondents' willingness to listen by 0.005 points, and increases, also non-significantly, by around one percent the length of audios they listen to. These results indicate that the high levels of engagement are not driven by the treatments. Even with no other information about the other person, participants show a remarkably high willingness to engage.  

It is possible that these average null treatment effects mask underlying heterogeneity. We examine this possibility by replicating the regressions by policy and by considering the initial position of the respondents' views on each policy. These additional analyses were not pre-registered and are therefore exploratory. See the results in \Autoref{t:wte_bypolicy} and \autoref{t:wte_bypreviews}. The regressions splitting by policy topic confirm the precisely null effects. The regressions examining each policy topic and conditioning on respondents' initial stand on such policies instead shows significant heterogeneous effects. 
Specifically, emphasizing shared support for behavioral etiquette rules increases a person's WTE on abortion and gun laws among people who had more conservative views on these policies. However, we also find that having more conservative views on abortion and gun laws are correlated with stronger preferences for behavioral etiquette rules; this is instead not the case for immigration, and this is also where we do not find an effect. Thus, it could be that agreeing with the basic etiquette rules is correlated with more conservative attitudes, and for this reason emphasizing common agreement on these rules might affect conservatives more.

\subsection{Changes in views}

The second outcome of interest is the change in views following the listening of the recordings. We are interested both in the magnitude of the change and the direction (towards more or less extreme values). 

\textbf{Result 3. Ten percent of participants report changing their views after listening to the recordings.}\\
The vast majority of participants does not change their view after having had the opportunity to listen to the recordings. Remarkably though, approximately 10\% of the respondents do report changing their views. This is true across all three policies. Given the short duration of the recordings, one would not expect views on such fundamental and polarizing policies to shift so rapidly, indicating that listening to others even for a short time might be a powerful intervention. 

Respondents who reported not having changed their views, were asked to explain why by selecting one of the following three options: "\textit{I already knew and considered the arguments}", "\textit{I didn't find the arguments convincing}", or "\textit{Other}". We see roughly the same split between the two main reasons across all three policies. About 51\% of those who didn't change their views on gun ownership laws said that they already knew the arguments, and 44\% said they didn't find the arguments convincing. On immigration laws, 48\% said they already considered the arguments, while 46\% didn't find them convincing. Lastly, about 52\% of those who didn't change their views on abortion laws said that they already knew the arguments, and 40\% said they didn't find the arguments convincing.

\begin{figure}[H]
     \centering
     \begin{subfigure}[b]{0.55\textwidth}
         \centering
         \includegraphics[width=\textwidth]{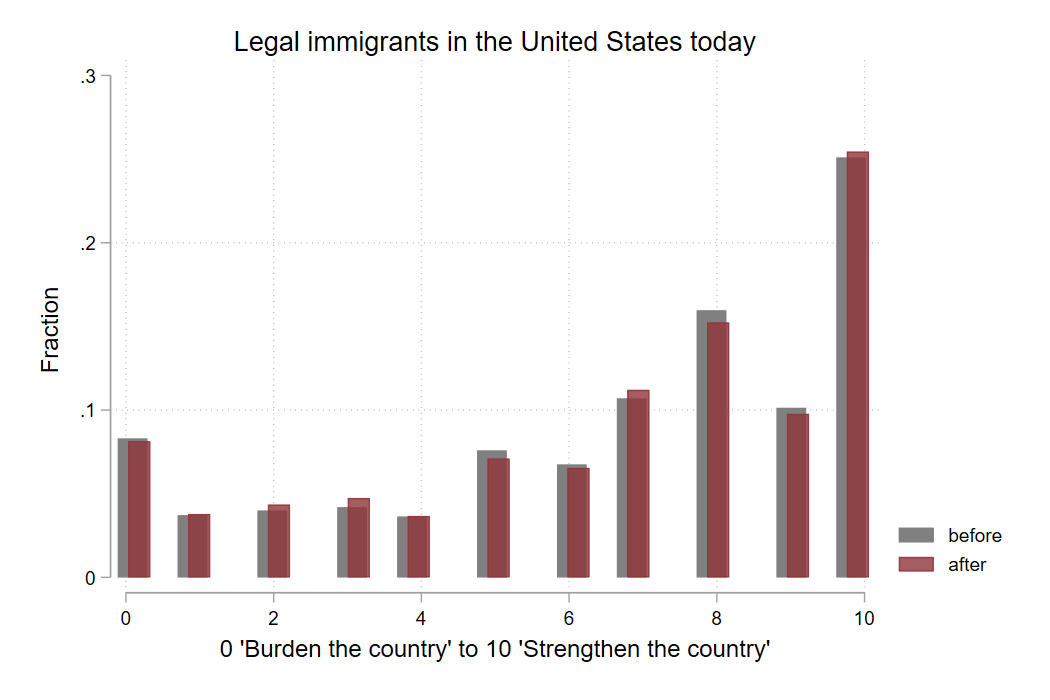}
         \label{fig:immi_changes}
     \end{subfigure}
     \hfill
     \begin{subfigure}[b]{0.55\textwidth}
         \centering
         \includegraphics[width=\textwidth]{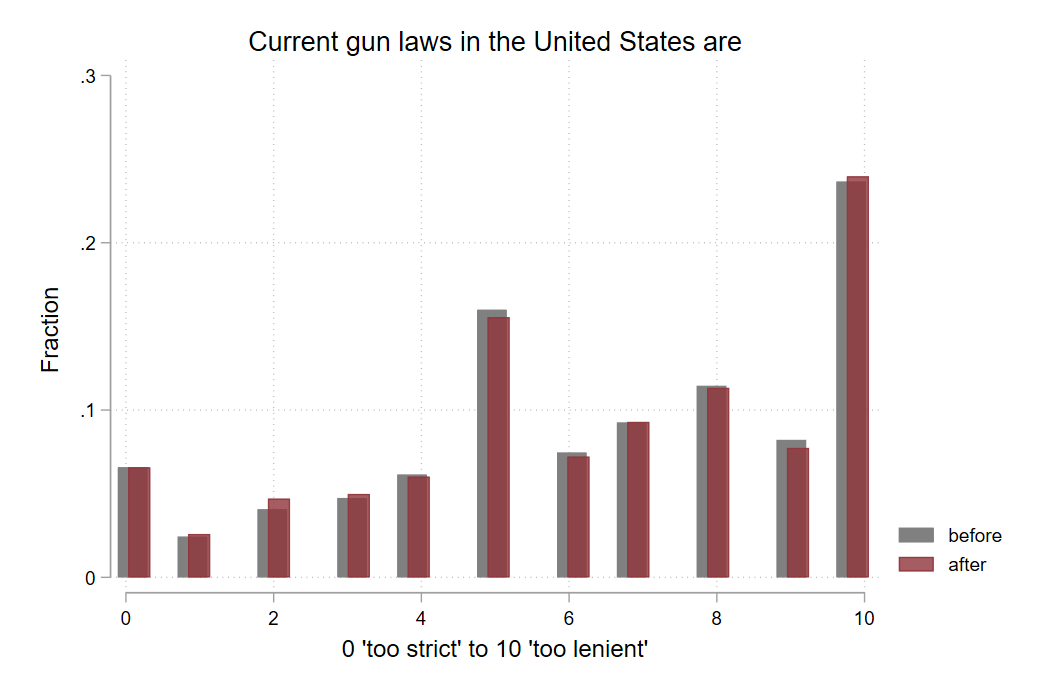}
         \label{fig:guns_changes}
     \end{subfigure}
     \hfill
     \begin{subfigure}[b]{0.55\textwidth}
         \centering
         \includegraphics[width=\textwidth]{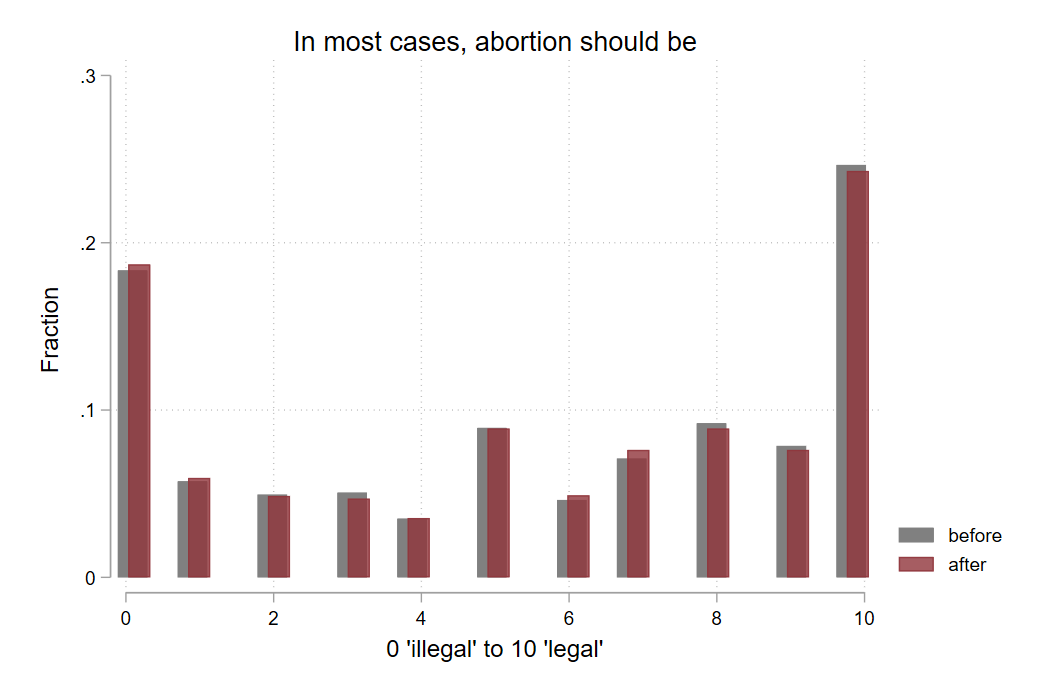}
         \label{fig:abort_changes}
     \end{subfigure}
        \caption{Changes in views}  \label{fig:all_changes}
        \label{fig:policy_changes}
  \vspace{-10pt}
   %\floatfoot{\textbf{\textit{Notes.}} The figure plots the distribution of before and after views on each policy.}
\end{figure}

\textbf{Result 4. In the aggregate, the distribution of views remains unchanged after listening to the recordings.}\\
\Autoref{t:views_before_after} and \Autoref{t:abs_views_before_after} present summary statistics of the views and their firmness (absolute distance from 5) for each of the policy. \Autoref{fig:all_changes} shows histograms of the distribution of views on each topic before and after having the opportunity to engage. We see that the distributions remain almost identical before and after. Among those who change view, the size of the change varies from 2 points on the scale from 1 to 10 (for the control group and gun laws) to 2.7 (for the Etiquette treatment and immigration). These changes are sizable and correspond to two-thirds of the standard deviation in the initial views.

%\Autoref{fig:policy_changes} plots the average difference between the views expressed after having the opportunity to listen to the recordings and the views expressed before, depending on these initial views. Those who responded that the audio recordings did not affect their views are given a value of zero. We find that for each topic, views tend to become less polarized after having had the opportunity to listen to the recordings. That is, overall, offering an opportunity to engage reduces polarization. 

\textbf{Result 5. Emphasizing common grounds does not affect views on average.}\\
We now turn to the treatment effects of emphasizing common grounds.  We pre-registered the analysis of how treatments might change views on the selected policies as a simple before-after comparison. 

In the first row of \Autoref{t:all_changes}, we report the average and standard deviation of each outcome variable. The next rows show the estimated coefficients. Here again, we do not find significant effects, but the estimates are less precise. From models (1) to (3) in \Autoref{t:all_changes} we can see that the average changes in views is about 0.35 for each policy. Compared to the control group, the T1 behavioral etiquette rules treatment increases changes views on immigration and abortion by a small and insignificant 0.01 points approximately, and it decreases it by around 0.03 points, also insignificantly, on views about gun ownership. The T2 human rights values treatment has a slightly larger effect on views on immigration and gun laws, and a very small negative effect on views about abortion laws, although all coefficients are statistically insignificant.  

Thus, so far, we find evidence that engagement is large and that a significant fraction of participants changes their views after having listened to others' views. In view of these very high baseline levels, it is perhaps not surprising that the treatment emphasizing common grounds does not have much effect.

As per the previous outcomes, we conduct additional heterogeneous analysis considering the respondents' initial views. We see that emphasizing shared beliefs on human rights values has an impact on views on gun laws for those who started with a more conservative initial position, but all the other estimated coefficients are not statistically significant (see \Autoref{t:change_views_2} for the regression results). 

\textbf{Result 6. Emphasizing common grounds leads to less polarized views on abortion and immigration.}\\
The last pre-registered outcome of interest is individuals' firmness of views -- that is, how polarized to the extremes individuals are on a scale 0 to 10 for each policy. 

Recall that the firmness of views is calculated as the absolute distance from 5 (the middle point on our 0 to 10 scale) for each policy. The difference in firmness before and after takes values of -5 to 5. 

We report the results in models (5) to (7) in \Autoref{t:all_changes}. Looking at the constant term first, we find that polarization increases marginally in the control group for views on abortion. There is no significant change in firmness of views on the two other topics. We also find compelling evidence that the treatments have an impact in reducing respondents' firmness of views. Knowing that the person who recorded the audio files shared the same views on behavioral etiquette rules (T1 treatment) significantly decreases the listeners' firmness of their views on immigration and abortion, although not on gun ownership laws. Knowing that the person who recorded the audio files shared the same human rights values (T2 treatment) significantly decreases the listeners' firmness of views on abortion. We find no significant treatment effects for views on gun laws.  

We also conduct a heterogeneity analysis, this time distinguishing between people who have more extreme values to start with (i.e. 8 or above, or 2 and below) from people who are more neutral (initial views between 3 and 7). We find that the treatment effects are driven by those who are more extreme on abortion laws, and become significantly less extreme when common grounds are emphasized.

\section{Conclusions}\label{sec:conclusion}

Political polarization is a major concern in American and other Western societies. In a world where citizens with different views do not talk and listen to one another online or offline, the opportunity to bridge gaps in views on important policies appears to be a challenging goal. A major obstacle in increasing contact between partisans is that people choose whom they interact with and could deliberately avoid engaging with others who think differently. Numerous efforts have been undertaken to increase contact between individuals with different backgrounds and political affinities, but these initiatives may not be very effective on a larger scale if people are not willing to listen to others who do not share their views without any incentive to do so. 

In this paper, we present evidence based on a representative sample of 2,507 respondents that Americans' willingness to engage with others on polarizing policies (abortion laws, gun ownership laws and immigration) is in fact high. We also find that a non-negligible fraction of people (~10 percent) reports changing their views after having listened to someone expressing their opposite views. We examined whether emphasizing common grounds, such as shared agreement on basic human rights or on simple behavioral etiquette rules, further raises the willingness to engage. This is not the case overall, although we have suggestive evidence that for more conservative Americans a shared view on behavioral etiquette rules might increase their willingness to listen to someone they disagree with. We find that listening to others who hold opposite views doesn't change the distribution of views on average but emphasizing common grounds does contribute to reduce polarization in views on abortion laws and immigration.

The paper is the first to examine directly the willingness to engage with others who do not share the same views on important policies. We also confirm Allport's hypothesis that contact is effective in altering views when common grounds are emphasized, both when these are common views on human rights and on more futile principles, such as basic behavioral etiquette rules. 

Overall, our take-away message is an encouraging one for those advocating a reduction in political polarization: Most people are willing to listen to others with opposite views, and a small but significant fraction of people is willing to change their views after listening to others they disagree with. Most importantly, overall views become less polarized after listening to others when common grounds are emphasized. These findings suggest that interventions aimed at bridging the partisan gap can be effective, particularly those that offer opportunities to voluntarily engage with others and emphasize common grounds.

\end{spacing}

\bibliographystyle{chicago}
\bibliography{openness}

\newpage

\section{Tables and Figures}

\vspace{0.8 cm}
%%%%%%%%%%%%%%%%%%%%%%%%%%%%%%%%%%%%%%%%%%%%%%%%%%%%%%%%%%%%%%%%%%%%%%%%%%%%%
\FloatBarrier
%%%%%%%%%%%%%%%%%%%%%%%%%%%%%%%%%%%%%%%%%%%%%%%%%%%%%%%%%%%%%%%%%%%%%%%%%%%%% 
\vspace{0.3 cm}

\vspace{0.8 cm}
%%%%%%%%%%%%%%%%%%%%%%%%%%%%%%%%%%%%%%%%%%%%%%%%%%%%%%%%%%%%%%%%%%%%%%%%%%%%%
\FloatBarrier
%%%%%%%%%%%%%%%%%%%%%%%%%%%%%%%%%%%%%%%%%%%%%%%%%%%%%%%%%%%%%%%%%%%%%%%%%%%%% 
\vspace{0.3 cm}

\vspace{0.8 cm}
%%%%%%%%%%%%%%%%%%%%%%%%%%%%%%%%%%%%%%%%%%%%%%%%%%%%%%%%%%%%%%%%%%%%%%%%%%%%%
\FloatBarrier
%%%%%%%%%%%%%%%%%%%%%%%%%%%%%%%%%%%%%%%%%%%%%%%%%%%%%%%%%%%%%%%%%%%%%%%%%%%%% 
\vspace{0.3 cm}

\vspace{0.8 cm}
%%%%%%%%%%%%%%%%%%%%%%%%%%%%%%%%%%%%%%%%%%%%%%%%%%%%%%%%%%%%%%%%%%%%%%%%%%%%%
\FloatBarrier
%%%%%%%%%%%%%%%%%%%%%%%%%%%%%%%%%%%%%%%%%%%%%%%%%%%%%%%%%%%%%%%%%%%%%%%%%%%%% 
\vspace{0.3 cm}

\vspace{0.6 cm}
%%%%%%%%%%%%%%%%%%%%%%%%%%%%%%%%%%%%%%%%%%%%%%%%%%%%%%%%%%%%%%%%%%%%%%%%%%%%%
\FloatBarrier
\begin{table}[H]
\begin{center}
\caption{WTE outcomes}
\label{t:wte}
\begin{tabular}{lcc}
\hline
               & (1)      & (2)      \\
 & \begin{tabular}[c]{@{}c@{}}Willingness \\ to engage\end{tabular} & \begin{tabular}[c]{@{}c@{}}Percentage total \\ audios length\\ listened to\end{tabular} \\ \hline
               &          &          \\
T1/etiq        & 0.00709  & 0.0120  \\
               & (0.0308) & (0.0335) \\
T2/values      & -0.00517 & 0.00136  \\
               & (0.0308) & (0.0333) \\
ln$\alpha$        & -36.07   &          \\
               &       &          \\
var(e.outcome) &          & 0.365*** \\
               &          & (0.0178) \\
Constant       & 0.925*** & 1.068*** \\
               & (0.0217) & (0.0250) \\
Observations   & 2,507    & 2,507   \\ \hline
\end{tabular}
\end{center}
\vspace{-25pt}
\floatfoot{\textbf{\textit{Notes.}} Model (1) is a negative binomial regression, with likelihood-ratio test of $\alpha$=0 is chibar2(01) = 0.00. Model (2) is a Tobit regression T1/etiq and T2/values refer to the two treatment groups. The pseudo-R$^2$ is less than 0 for both models. {$^{***}\ p<0.01$; $^{**}\ p<0.05$; $^{*}\ p<0.1$}.}
\end{table}
%%%%%%%%%%%%%%%%%%%%%%%%%%%%%%%%%%%%%%%%%%%%%%%%%%%%%%%%%%%%%%%%%%%%%%%%%%%%%

\begin{landscape}

\vspace{0.6 cm}
%%%%%%%%%%%%%%%%%%%%%%%%%%%%%%%%%%%%%%%%%%%%%%%%%%%%%%%%%%%%%%%%%%%%%%%%%%%%%
\FloatBarrier
\begin{table}[H]
\begin{center}
\captionsetup{width=0.9\columnwidth,labelfont=bf}
\caption{Views and Firmness of Views - Before and After Opportunity to Engage}
\label{t:views_before_after}
\begin{tabular}{lccccccc}
\cline{2-4} \cline{6-8}
 & \multicolumn{3}{c}{\begin{tabular}[c]{@{}c@{}}Views (0-10)\\      Means (standard deviations)\end{tabular}} & \multicolumn{1}{l}{} & \multicolumn{3}{c}{\begin{tabular}[c]{@{}c@{}}Firmness of   views\\      (distance from 5)\\      Means\end{tabular}} \\ \cline{2-4} \cline{6-8} 
 & \multicolumn{1}{l}{Immigration} & \multicolumn{1}{l}{Gun laws} & \multicolumn{1}{l}{Abortion} & \multicolumn{1}{l}{} & \multicolumn{1}{l}{Immigration} & \multicolumn{1}{l}{Gun laws} & \multicolumn{1}{l}{Abortion} \\ \cline{1-4} \cline{6-8} 
 & \multicolumn{1}{l}{} & \multicolumn{1}{l}{} & \multicolumn{1}{l}{} & \multicolumn{1}{l}{} & \multicolumn{1}{l}{} & \multicolumn{1}{l}{} & \multicolumn{1}{l}{} \\
All - before & 6.6 (3.2) & 6.4 (3.1) & 5.6 (3.8) &  & 3.2 & 2.8 & 3.5 \\
All - after & 6.6 (3.2) & 6.4 (3.1) & 5.5 (3.8) &  & 3.2 & 2.8 & 3.5 \\
\textit{p-value, t-test before= after} & 0.77 & 0.66 & 0.7 &  & 0.00 & 0.69 & 0.77 \\ \cline{1-4} \cline{6-8} 
 &  &  &  &  &  &  &  \\
(1) Control - before & 6.6 (3.2) & 6.4 (3.1) & 5.5 (3.7) &  & 3.2 & 2.8 & 3.3 \\
(2) Control - after & 6.6 (3.2) & 6.3 (3.1) & 5.4 (3.8) &  & 3.2 & 2.8 & 3.4 \\
\textit{p-value, t-test (1)=(2)} & 0.95 & 0.45 & 0.70 &  & 0.69 & 0.94 & 0.63 \\ \cline{1-4} \cline{6-8} 
 & \multicolumn{1}{l}{} & \multicolumn{1}{l}{} & \multicolumn{1}{l}{} & \multicolumn{1}{l}{} & \multicolumn{1}{l}{} & \multicolumn{1}{l}{} & \multicolumn{1}{l}{} \\
(3) T1/etiq - before & 6.5 (3.3) & 6.4 (3.1) & 5.6 (3.8) &  & 3.3 & 2.8 & 3.5 \\
(4) T1/etiq - after & 6.4 (3.3) & 6.4 (3.1) & 5.6 (3.8) &  & 3.1 & 2.9 & 3.4 \\
\textit{p-value, t-test (3)=(4)} & 0.51 & 0.96 & 0.86 &  & 0.74 & 0.69 & 0.65 \\ \cline{1-4} \cline{6-8} 
 &  &  &  &  &  &  &  \\
(5) T2/values - before & 6.8 (3.2) & 6.4 (3.1) & 5.6 (3.8) &  & 3.3 & 2.9 & 3.5 \\
(6) T2/values - after & 6.8 (3.3) & 6.4 (3.1) & 5.6 (3.8) &  & 3.3 & 2.9 & 3.5 \\
\textit{p-value, t-test (5)=(6)} & 0.93 & 0.98 & 0.67 &  & 0.81 & 0.83 & 0.77 \\ \cline{1-4} \cline{6-8} 
 &  &  &  &  &  &  &  \\
\textit{p-value, DD T1/etiq vs Control} & 0.09 & 0.12 & 0.69 &  & 0.06 & 0.32 & 0.01 \\
\textit{p-value, DD T2/values vs  Control} & 0.94 & 0.08 & 0.27 &  & 0.70 & 0.68 & 0.03 \\ \cline{1-4} \cline{6-8} 
\end{tabular}
\end{center}
\vspace{-25pt}
\floatfoot{\textbf{\textit{Notes.}} The table reports the average and standard deviation (in parenthesis) for views and firmness of views (distance from 5) before and after the opportunity to listen to the recordings.}
\end{table}
%%%%%%%%%%%%%%%%%%%%%%%%%%%%%%%%%%%%%%%%%%%%%%%%%%%%%%%%%%%%%%%%%%%%%%%%%%%%%

\vspace{0.6 cm}
%%%%%%%%%%%%%%%%%%%%%%%%%%%%%%%%%%%%%%%%%%%%%%%%%%%%%%%%%%%%%%%%%%%%%%%%%%%%%
\FloatBarrier
\begin{table}[H]
\begin{center}
\captionsetup{width=0.9\columnwidth,labelfont=bf}
\caption{Change in Views (Absolute values) - Overall and Conditional on Changing Views}
\label{t:abs_views_before_after}
\begin{tabular}{lccccccc}
\cline{2-4} \cline{6-8}
 & \multicolumn{3}{c}{\begin{tabular}[c]{@{}c@{}}Absolute change in views \\       $|$after - before$|$\end{tabular}} & \multicolumn{1}{l}{} & \multicolumn{3}{c}{\begin{tabular}[c]{@{}c@{}}Absolute change in views \\      $|$after - before$|$\\ (conditional on changing)\end{tabular}} \\ \cline{2-4} \cline{6-8} 
 & Immigration & Gun laws & Abortion &  & Immigration & Gun laws & Abortion \\ \cline{1-4} \cline{6-8} 
All &  &  &  &  &  &  &  \\
(1) Control & 0.3 (1.3) & 0.3 (1.2) & 0.3 (1.3) &  & 2.3 (2.5) & 2.0 (2.3) & 2.6 (2.9) \\
(2) T1/etiq & 0.4 (1.4) & 0.3 (1.2) & 0.3 (1.3) &  & 2.7 (2.8) & 2.4 (2.7) & 2.6 (2.8) \\
(3) T2/values & 0.4 (1.4) & 0.4 (1.4) & 0.3 (1.3) &  & 2.4 (2.6) & 2.5 (2.7) & 2.4 (2.9) \\ \cline{1-4} \cline{6-8} 
 &  &  &  &  &  &  &  \\
\textit{p-value, t-test (1)=(2)} & 0.84 & 0.53 & 0.87 &  & 0.24 & 0.29 & 0.97 \\
\textit{p-value, t-test (1)=(3)} & 0.9 & 0.33 & 0.43 &  & 0.66 & 0.16 & 0.51 \\ \cline{1-4} \cline{6-8} 
\end{tabular}
\end{center}
\vspace{-25pt}
\floatfoot{\textbf{\textit{Notes.}} The table reports the average and standard deviation (in parenthesis) for absolute changes in views following the opportunity to listen to the recordings.}
\end{table}
%%%%%%%%%%%%%%%%%%%%%%%%%%%%%%%%%%%%%%%%%%%%%%%%%%%%%%%%%%%%%%%%%%%%%%%%%%%%%
\end{landscape}

\newpage

\begin{landscape}

\vspace{0.6 cm}
%%%%%%%%%%%%%%%%%%%%%%%%%%%%%%%%%%%%%%%%%%%%%%%%%%%%%%%%%%%%%%%%%%%%%%%%%%%%%
\FloatBarrier
\begin{table}[H]
\begin{center}
\caption{Changes in views and firmness of views}
\label{t:all_changes}
\begin{tabular}{lcccccc}
\hline
               & (1)                  & (2)                  & (3)                  & (4)               & (5)               & (6)               \\
 &
  \begin{tabular}[c]{@{}c@{}}$\Delta$ in views\\ Immigration\end{tabular} &
  \begin{tabular}[c]{@{}c@{}}$\Delta$ in views\\ Abortion\end{tabular} &
  \begin{tabular}[c]{@{}c@{}}$\Delta$ in views\\ Gun laws\end{tabular} &
  \begin{tabular}[c]{@{}c@{}}$\Delta$ in firmn.\\ of views\\ Immigration\end{tabular} &
  \begin{tabular}[c]{@{}c@{}}$\Delta$ in firmn.\\ of views\\ Abortion\end{tabular} &
  \begin{tabular}[c]{@{}c@{}}$\Delta$ in firmn.\\ of views\\ Gun laws\end{tabular} \\ \hline
\textit{Avg}   & \textit{0.362}       & \textit{0.329}       & \textit{0.347}       & \textit{0.008}    & \textit{- 0.0071} & \textit{0.0207}   \\
\textit{}      & \textit{(1.337)}     & \textit{(1.345)}     & \textit{(1.301)}     & \textit{(0.6214)} & \textit{(0.6188)} & \textit{(0.6002)} \\
T1/etiq        & 0.0110               & 0.0132               & -0.0398              & -0.0575*          & -0.0791***        & 0.0291            \\
               & (0.0655)             & (0.0659)             & (0.0637)             & (0.0304)          & (0.0303)          & (0.0294)          \\
T2/values      & 0.0518               & -0.00798             & 0.0613               & -0.0118           & -0.0642**         & 0.0120            \\
               & (0.0653)             & (0.0657)             & (0.0635)             & (0.0303)          & (0.0302)          & (0.0293)          \\
Constant       & 0.342***             & 0.328***             & 0.340***             & 0.0309            & 0.0404*           & 0.00713           \\
               & (0.0461)             & (0.0464)             & (0.0449)             & (0.0214)          & (0.0213)          & (0.0207)          \\
               & \multicolumn{1}{l}{} & \multicolumn{1}{l}{} & \multicolumn{1}{l}{} &                   &                   &                   \\
Observations   & 2,507                & 2,507                & 2,507                & 2,507             & 2,507             & 2,507             \\
\textit{R$^2$} & 0.000                & 0.000                & 0.001                & 0.002             & 0.003             & 0.000             \\ \hline
\end{tabular}
\end{center}
\vspace{-25pt}
\floatfoot{\textbf{\textit{Notes.}} Models (1) to (3) report changes in views, while models (4) to (6) report changes in firmness of views. The first row reports the average and standard deviation of each outcome variable, which can take values between -5 and +5 for models (4) to (6). All models are OLS.{$^{***}\ p<0.01$; $^{**}\ p<0.05$; $^{*}\ p<0.1$}.}
\end{table}
%%%%%%%%%%%%%%%%%%%%%%%%%%%%%%%%%%%%%%%%%%%%%%%%%%%%%%%%%%%%%%%%%%%%%%%%%%%%%

\end{landscape}

\newpage

\appendix

\setcounter{table}{0}

\section{Appendix A}

\subsection{Additional tables and figures} \label{sec:Appendix_tables}

\renewcommand{\thetable}{A\arabic{table}}

\vspace{0.6 cm}
%%%%%%%%%%%%%%%%%%%%%%%%%%%%%%%%%%%%%%%%%%%%%%%%%%%%%%%%%%%%%%%%%%%%%%%%%%%%%
\FloatBarrier
\begin{table}[H]
\begin{center}
\caption{Sample characteristics \& balance checks}
\label{t:balance}
\begin{tabular}{l*{6}c}
\hline\hline
 & (1) & (2) & (3) & (4) & (5) & (6) \\
 & Control & T1/etiquette & T2/values & T1 vs C & T2 vs C & T1 vs T2 \\
\hline
Female&0.499&0.475&0.484&-0.024&-0.014&-0.009\\
&(0.500)&(0.500)&(0.500)&(0.024)&(0.024)&(0.025)\\
Age 18-34&0.305&0.276&0.261&-0.030&-0.044**&0.014\\
&(0.461)&(0.447)&(0.440)&(0.022)&(0.022)&(0.022)\\
Age 55-70&0.297&0.289&0.310&-0.008&0.013&-0.021\\
&(0.457)&(0.454)&(0.463)&(0.022)&(0.022)&(0.022)\\
Age 35-54&0.279&0.307&0.297&0.028&0.018&0.010\\
&(0.449)&(0.462)&(0.457)&(0.022)&(0.022)&(0.023)\\
Age 70+&0.118&0.126&0.130&0.008&0.012&-0.004\\
&(0.322)&(0.332)&(0.337)&(0.016)&(0.016)&(0.016)\\
HS educ or more&0.982&0.985&0.971&0.003&-0.011&0.014**\\
&(0.132)&(0.120)&(0.167)&(0.006)&(0.007)&(0.007)\\
Income 31K or less&0.385&0.393&0.391&0.008&0.007&0.002\\
&(0.487)&(0.489)&(0.488)&(0.024)&(0.024)&(0.024)\\
Income 31-70K&0.420&0.393&0.407&-0.027&-0.014&-0.014\\
&(0.494)&(0.489)&(0.492)&(0.024)&(0.024)&(0.024)\\
Income 70K+&0.195&0.214&0.202&0.019&0.007&0.012\\
&(0.396)&(0.410)&(0.401)&(0.020)&(0.019)&(0.020)\\
Democrat&0.398&0.366&0.405&-0.031&0.007&-0.038\\
&(0.490)&(0.482)&(0.491)&(0.024)&(0.024)&(0.024)\\
Republican&0.397&0.417&0.407&0.020&0.010&0.010\\
&(0.489)&(0.493)&(0.492)&(0.024)&(0.024)&(0.024)\\
Indep/Centrist/others&0.205&0.216&0.189&0.011&-0.017&0.028\\
&(0.404)&(0.412)&(0.391)&(0.020)&(0.019)&(0.020)\\
\hline
Observations & 842 & 827 & 838 & 1,669 & 1,680 & 1,665 \\
\hline\hline
\end{tabular}
\end{center}
\vspace{-25pt}
\floatfoot{\textbf{\textit{Notes.}} Columns 1 to 3 report the (mean) share of respondents in each of the three experimental groups across key demographic and socio-economic characteristics, with standard deviation in parenthesis. Columns 3 to 6 report the results of between sub-sample balance tests. We record a marginally smaller share (4\%) of respondents aged 18 to 34 in the T2 group compared to control, and a marginally larger share (1\%) of more educated respondents in T1 compared to T2. None of these differences affect our main results in any meaningful way. {$^{***}\ p<0.01$; $^{**}\ p<0.05$; $^{*}\ p<0.1$}.}
\end{table}
%%%%%%%%%%%%%%%%%%%%%%%%%%%%%%%%%%%%%%%%%%%%%%%%%%%%%%%%%%%%%%%%%%%%%%%%%%%%%

\renewcommand{\thetable}{A\arabic{table}}

\vspace{0.6 cm}
%%%%%%%%%%%%%%%%%%%%%%%%%%%%%%%%%%%%%%%%%%%%%%%%%%%%%%%%%%%%%%%%%%%%%%%%%%%%%
\FloatBarrier
\begin{table}[H]
\begin{center}
\caption{Engagement Levels - Main Survey VS Baseline Comparison Survey}
\label{t:wte_baseline}
\begin{tabular}{lcc}
\cline{2-3}
 & \begin{tabular}[c]{@{}c@{}}Main\\ Survey\end{tabular} & \begin{tabular}[c]{@{}c@{}}Baseline\\ Survey\end{tabular} \\ \hline
Share of respondents who listened to all three files & 69.5\% & 69.1\% \\
Share of respondents who listened to two files & 18.3\% & 18.2\% \\
\begin{tabular}[c]{@{}l@{}}Share of respondents who listened to at least the \\ total length of audio files\end{tabular} & 56.3\% & 54.5\% \\
Age 18-34 & 28\% & 18\% \\
Age 35-54 & 29\% & 54\% \\
Age 55-70 & 30\% & 22\% \\
Age 71+ & 13\% & 6\% \\
HS educ or more & 98\% & 98\% \\
Income 31K or less & 39\% & 40\% \\
Income 31-70K & 41\% & 35\% \\
Income 70K+ & 20\% & 25\% \\
Democrat & 39\% & 40\% \\
Republican & 41\% & 40\% \\
Indep/Centrist/others & 20\% & 20\% \\ \hline
Observations & 2,507 & 110 \\ \hline
\end{tabular}
\end{center}
\vspace{-25pt}
\floatfoot{\textbf{\textit{Notes.}} The Table shows the share of respondents within each survey - main survey and the additional baseline comparison survey - against the pre-registered outcome variables and demographic characteristics.}
\end{table}
%%%%%%%%%%%%%%%%%%%%%%%%%%%%%%%%%%%%%%%%%%%%%%%%%%%%%%%%%%%%%%%%%%%%%%%%%%%%%

\vspace{0.8 cm}
%%%%%%%%%%%%%%%%%%%%%%%%%%%%%%%%%%%%%%%%%%%%%%%%%%%%%%%%%%%%%%%%%%%%%%%%%%%%%
\FloatBarrier
\begin{figure}[H]
   \centering
    \caption{Time spent on page as fraction of length of recordings}
    \label{fig:tot_perc_engage}
   \includegraphics[height=8cm,angle=0]{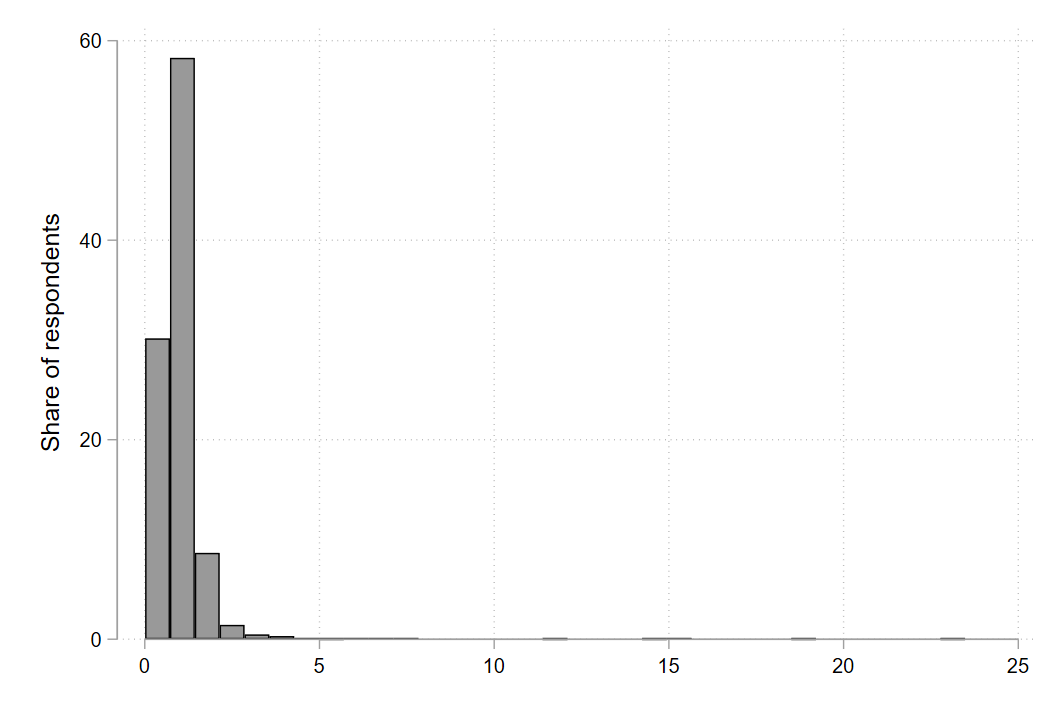}
   \vspace{-10pt}
   \floatfoot{\textbf{\textit{Notes.}} The figure plots the frequency of the time spent on the three audio files as a fraction of the total length of the recordings.}
  \end{figure}
%%%%%%%%%%%%%%%%%%%%%%%%%%%%%%%%%%%%%%%%%%%%%%%%%%%%%%%%%%%%%%%%%%%%%%%%%%%%% 
\vspace{0.3 cm}

\vspace{0.8 cm}
%%%%%%%%%%%%%%%%%%%%%%%%%%%%%%%%%%%%%%%%%%%%%%%%%%%%%%%%%%%%%%%%%%%%%%%%%%%%%
\FloatBarrier
\begin{figure}[H]
   \centering
    \caption{Time spent on page as fraction of length of recordings, censored}
    \label{fig:tot_perc_engage}
   \includegraphics[height=8cm,angle=0]{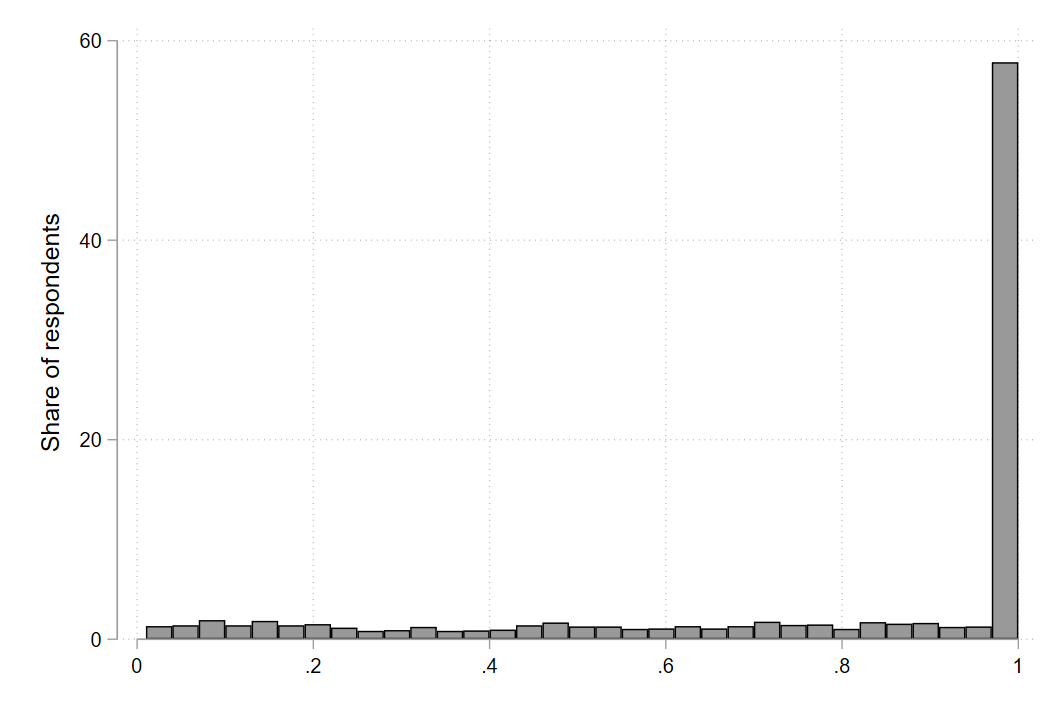}
   \vspace{-10pt}
   \floatfoot{\textbf{\textit{Notes.}} The figure plots the frequency of the time spent on the three audio files as a fraction of the total length of the recordings. Any values above 1 is equalized to 1, as per Tobit regressions in the main text.}
  \end{figure}
%%%%%%%%%%%%%%%%%%%%%%%%%%%%%%%%%%%%%%%%%%%%%%%%%%%%%%%%%%%%%%%%%%%%%%%%%%%%% 
\vspace{0.3 cm}

\vspace{0.8 cm}
%%%%%%%%%%%%%%%%%%%%%%%%%%%%%%%%%%%%%%%%%%%%%%%%%%%%%%%%%%%%%%%%%%%%%%%%%%%%%
\FloatBarrier
\begin{figure}[H]
   \centering
    \caption{Self-reported engagement}
    \label{fig:self_reported_eng}
   \includegraphics[height=8cm,angle=0]{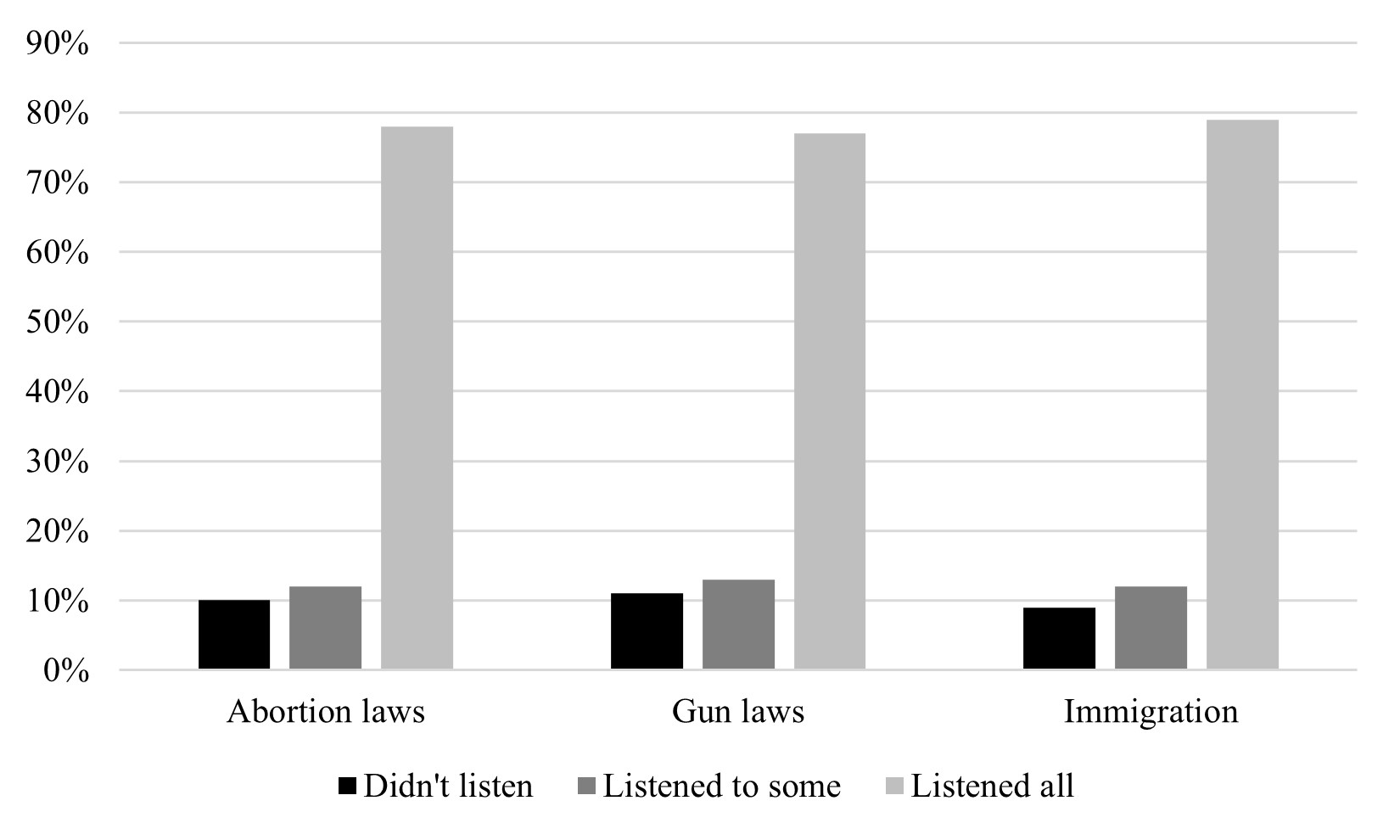}
   \vspace{-10pt}
   \floatfoot{\textbf{\textit{Notes.}} The figure plots the self-reported engagement with the recordings for each policy.}
  \end{figure}
%%%%%%%%%%%%%%%%%%%%%%%%%%%%%%%%%%%%%%%%%%%%%%%%%%%%%%%%%%%%%%%%%%%%%%%%%%%%% 
\vspace{0.3 cm}

\vspace{0.6 cm}
%%%%%%%%%%%%%%%%%%%%%%%%%%%%%%%%%%%%%%%%%%%%%%%%%%%%%%%%%%%%%%%%%%%%%%%%%%%%%
\FloatBarrier
\begin{table}[H]
\begin{center}
\caption{WTE by policy}
\label{t:wte_bypolicy}
\begin{tabular}{lccc}
\hline
             & (1)      & (2)      & (3)      \\
 &
  \begin{tabular}[c]{@{}c@{}}WTE \\ Immigration\end{tabular} &
  \begin{tabular}[c]{@{}c@{}}WTE \\ Abortion\end{tabular} &
  \begin{tabular}[c]{@{}c@{}}WTE \\ Gun laws\end{tabular} \\ \hline
             &          &          &          \\
T1/etiq      & -0.121   & 0.131    & 0.0706   \\
             & (0.159)  & (0.130)  & (0.124)  \\
T2/values    & -0.106   & 0.0268   & -0.0431  \\
             & (0.159)  & (0.128)  & (0.122)  \\
Constant     & 2.187*** & 1.521*** & 1.404*** \\
             & (0.114)  & (0.0898) & (0.0866) \\
             &          &          &          \\
Observations & 2,507    & 2,507    & 2,507    \\ \hline
\end{tabular}%
\end{center}
\vspace{-25pt}
\floatfoot{\textbf{\textit{Notes.}} Models (1) to (3) are Logit regressions. T1/etiq and T2/values refer to the two treatment groups.{$^{***}\ p<0.01$; $^{**}\ p<0.05$; $^{*}\ p<0.1$}.}
\end{table}
%%%%%%%%%%%%%%%%%%%%%%%%%%%%%%%%%%%%%%%%%%%%%%%%%%%%%%%%%%%%%%%%%%%%%%%%%%%%%

\vspace{0.6 cm}
%%%%%%%%%%%%%%%%%%%%%%%%%%%%%%%%%%%%%%%%%%%%%%%%%%%%%%%%%%%%%%%%%%%%%%%%%%%%%
\FloatBarrier
\begin{table}[H]
\begin{center}
\caption{WTE by policy and pre-treatment views}
\label{t:wte_bypreviews}
\begin{tabular}{lcccccc}
\hline
             & (1)      & (2)      & (3)      & (4)      & (5)      & (6)      \\
 &
  \begin{tabular}[c]{@{}c@{}}WTE \\ Immigration\end{tabular} &
  \begin{tabular}[c]{@{}c@{}}WTE \\ Immigration\end{tabular} &
  \begin{tabular}[c]{@{}c@{}}WTE \\ Abortion\end{tabular} &
  \begin{tabular}[c]{@{}c@{}}WTE \\ Abortion\end{tabular} &
  \begin{tabular}[c]{@{}c@{}}WTE \\ Gun laws\end{tabular} &
  \begin{tabular}[c]{@{}c@{}}WTE \\ Gun laws\end{tabular} \\ \hline
             &          &          &          &          &          &          \\
T1/etiq      & 0.130    & -0.311   & 0.461**  & -0.0605  & 0.505*   & -0.0629  \\
             & (0.249)  & (0.213)  & (0.218)  & (0.165)  & (0.259)  & (0.142)  \\
T2/values    & -0.0293  & -0.214   & 0.112    & -0.0323  & 0.0965   & -0.0870  \\
             & (0.246)  & (0.215)  & (0.200)  & (0.166)  & (0.239)  & (0.142)  \\
Constant     & 1.326*** & 2.651*** & 1.379*** & 1.616*** & 1.191*** & 1.473*** \\
             & (0.172)  & (0.160)  & (0.139)  & (0.118)  & (0.170)  & (0.101)  \\
             &          &          &          &          &          &          \\
Observations & 597      & 1,910    & 943      & 1,564    & 601      & 1,906    \\ \hline
\end{tabular}
\end{center}
\vspace{-25pt}
\floatfoot{\textbf{\textit{Notes.}} All models are Logit regressions. Models (1), (3), and (5) only consider the subset of respondents who gave a score strictly less than 5 out of 10 on each of the policies at the beginning of the experiment. Models (2), (4) and (6) only consider the subset of respondents who gave a score 5 or above out of 10 on each of the policies at the beginning of the experiment. T1/etiq and T2/values refer to the two treatment groups.{$^{***}\ p<0.01$; $^{**}\ p<0.05$; $^{*}\ p<0.1$}.}
\end{table}
%%%%%%%%%%%%%%%%%%%%%%%%%%%%%%%%%%%%%%%%%%%%%%%%%%%%%%%%%%%%%%%%%%%%%%%%%%%%%

\vspace{0.6 cm}
%%%%%%%%%%%%%%%%%%%%%%%%%%%%%%%%%%%%%%%%%%%%%%%%%%%%%%%%%%%%%%%%%%%%%%%%%%%%%
\FloatBarrier
\begin{table}[H]
\begin{center}
\caption{Changes in views conditional on priors}
\label{t:change_views_2}
\begin{tabular}{lcccccc}
\hline
             & (1)      & (2)      & (3)      & (4)      & (5)      & (6)      \\
 &
  \begin{tabular}[c]{@{}c@{}}$\Delta$ in views\\ Immigration\end{tabular} &
  \begin{tabular}[c]{@{}c@{}}$\Delta$in views\\ Immigration\end{tabular} &
  \begin{tabular}[c]{@{}c@{}}$\Delta$ in views \\ Abortion\end{tabular} &
  \begin{tabular}[c]{@{}c@{}}$\Delta$ in views \\ Abortion\end{tabular} &
  \begin{tabular}[c]{@{}c@{}}$\Delta$ in views\\ Gun laws\end{tabular} &
  \begin{tabular}[c]{@{}c@{}}$\Delta$ in views\\ Gun laws\end{tabular} \\ \hline
             &          &          &          &          &          &          \\
T1/etiq      & -0.154   & 0.0641   & 0.0232   & 0.00771  & 0.0780   & -0.0774  \\
             & (0.174)  & (0.0662) & (0.107)  & (0.0836) & (0.170)  & (0.0644) \\
T2/values    & 0.174    & 0.0203   & -0.0571  & 0.0225   & 0.331*   & -0.0279  \\
             & (0.176)  & (0.0656) & (0.105)  & (0.0841) & (0.169)  & (0.0643) \\
Constant     & 0.522*** & 0.284*** & 0.331*** & 0.326*** & 0.342*** & 0.339*** \\
             & (0.122)  & (0.0466) & (0.0743) & (0.0594) & (0.121)  & (0.0451) \\
             &          &          &          &          &          &          \\
Observations & 597      & 1,910    & 943      & 1,564    & 601      & 1,906    \\
R-squared    & 0.006    & 0.001    & 0.001    & 0.000    & 0.007    & 0.001    \\ \hline
\end{tabular}
\end{center}
\vspace{-25pt}
\floatfoot{\textbf{\textit{Notes.}} All models are OLS regressions. Models (1), (3), and (5) only consider the subset of respondents who gave a score strictly less than 5 out of 10 on each of the policies at the beginning of the experiment. Models (2), (4) and (6) only consider the subset of respondents who gave a score 5 or above out of 10 on each of the policies at the beginning of the experiment. T1/etiq and T2/values refer to the two treatment groups.{$^{***}\ p<0.01$; $^{**}\ p<0.05$; $^{*}\ p<0.1$}.}
\end{table}
%%%%%%%%%%%%%%%%%%%%%%%%%%%%%%%%%%%%%%%%%%%%%%%%%%%%%%%%%%%%%%%%%%%%%%%%%%%%%

\vspace{0.6 cm}
%%%%%%%%%%%%%%%%%%%%%%%%%%%%%%%%%%%%%%%%%%%%%%%%%%%%%%%%%%%%%%%%%%%%%%%%%%%%%
\FloatBarrier
\begin{landscape}
\begin{table}[H]
\begin{center}
\caption{Changes in firmness of views conditional on priors}
\label{t:firmness_views2}
\begin{tabular}{lcccccc}
\hline
             & (1)      & (2)       & (3)      & (4)       & (5)      & (6)       \\
 &
  \begin{tabular}[c]{@{}c@{}}Changes firmn.\\ of views \\ Immigration\\ {[}Neutral{]}\end{tabular} &
  \begin{tabular}[c]{@{}c@{}}Changes firmn.\\ of views \\ Immigration\\ {[}Extreme{]}\end{tabular} &
  \begin{tabular}[c]{@{}c@{}}Changes firmn.\\ of views \\ Abortion\\ {[}Neutral{]}\end{tabular} &
  \begin{tabular}[c]{@{}c@{}}Changes firmn.\\ of views \\ Abortion\\ {[}Extreme{]}\end{tabular} &
  \begin{tabular}[c]{@{}c@{}}Changes firmn.\\ of views \\ Gun laws\\ {[}Neutral{]}\end{tabular} &
  \begin{tabular}[c]{@{}c@{}}Changes firmn.\\ of views \\ Gun laws\\ {[}Extreme{]}\end{tabular} \\ \hline
             &          &           &          &           &          &           \\
T1/etiq      & 0.117    & 0.0783    & -0.0482  & -0.195**  & 0.268    & -0.000787 \\
             & (0.156)  & (0.0934)  & (0.0300) & (0.0957)  & (0.173)  & (0.0994)  \\
T2/values    & 0.0549   & 0.0441    & -0.0131  & -0.265*** & -0.0277  & 0.148     \\
             & (0.157)  & (0.0923)  & (0.0297) & (0.0982)  & (0.175)  & (0.0983)  \\
Constant     & 1.577*** & -0.263*** & -0.0307  & 0.350***  & 2.336*** & -0.237*** \\
             & (0.110)  & (0.0654)  & (0.0212) & (0.0660)  & (0.123)  & (0.0696)  \\
             &          &           &          &           &          &           \\
Observations & 823      & 1,684     & 2,079    & 428       & 742      & 1,765     \\
R-squared    & 0.001    & 0.000     & 0.001    & 0.019     & 0.005    & 0.002     \\ \hline
\end{tabular}
\end{center}
\vspace{-25pt}
\floatfoot{\textbf{\textit{Notes.}} All models are OLS regressions. Models (1), (3), and (5) only consider the subset of respondents who gave a score between 3 and 7 (included) on each of the policies at the beginning of the experiment. Models (2), (4) and (6) only consider the subset of respondents who gave a score of 0 to 2, or 8 to 10 on each of the policies at the beginning of the experiment. T1/etiq and T2/values refer to the two treatment groups.{$^{***}\ p<0.01$; $^{**}\ p<0.05$; $^{*}\ p<0.1$}.}
\end{table}
\end{landscape}
%%%%%%%%%%%%%%%%%%%%%%%%%%%%%%%%%%%%%%%%%%%%%%%%%%%%%%%%%%%%%%%%%%%%%%%%%%%%%

\section{Appendix B}

\subsection{Listening instructions and audio files} \label{sec:Appendix_instructions}

\vspace{0.8 cm}
%%%%%%%%%%%%%%%%%%%%%%%%%%%%%%%%%%%%%%%%%%%%%%%%%%%%%%%%%%%%%%%%%%%%%%%%%%%%%
\FloatBarrier
%%%%%%%%%%%%%%%%%%%%%%%%%%%%%%%%%%%%%%%%%%%%%%%%%%%%%%%%%%%%%%%%%%%%%%%%%%%%% 
\vspace{0.3 cm}

\vspace{0.8 cm}
%%%%%%%%%%%%%%%%%%%%%%%%%%%%%%%%%%%%%%%%%%%%%%%%%%%%%%%%%%%%%%%%%%%%%%%%%%%%%
\FloatBarrier
\begin{figure}[H]
   \centering
    \caption{Audio file display page example - Main survey}
    \label{fig:instructions_audio2}
   \includegraphics[height=9.5cm,angle=0]{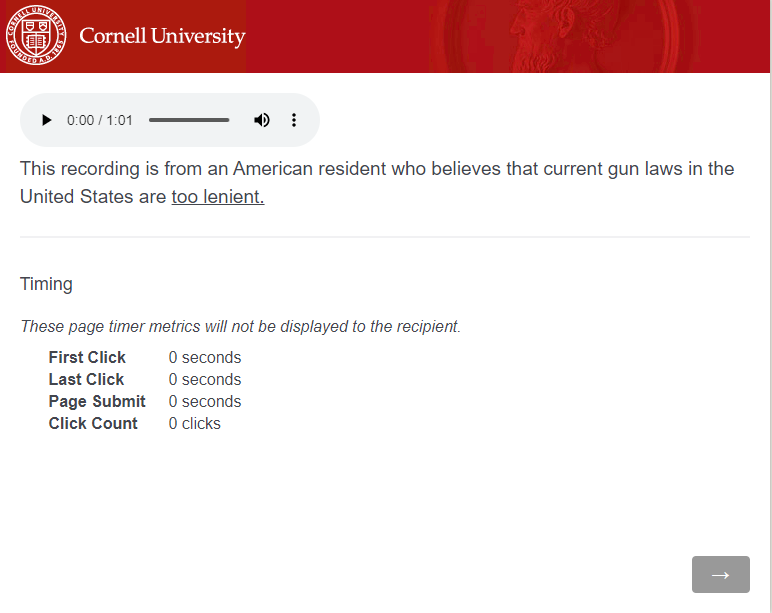}
  \end{figure}
%%%%%%%%%%%%%%%%%%%%%%%%%%%%%%%%%%%%%%%%%%%%%%%%%%%%%%%%%%%%%%%%%%%%%%%%%%%%% 
\vspace{0.3 cm}

\vspace{0.8 cm}
%%%%%%%%%%%%%%%%%%%%%%%%%%%%%%%%%%%%%%%%%%%%%%%%%%%%%%%%%%%%%%%%%%%%%%%%%%%%%
\FloatBarrier
\begin{figure}[H]
   \centering
    \caption{Audio file display page example - Baseline survey}
    \label{fig:instructions_audio2}
   \includegraphics[height=9.5cm,angle=0]{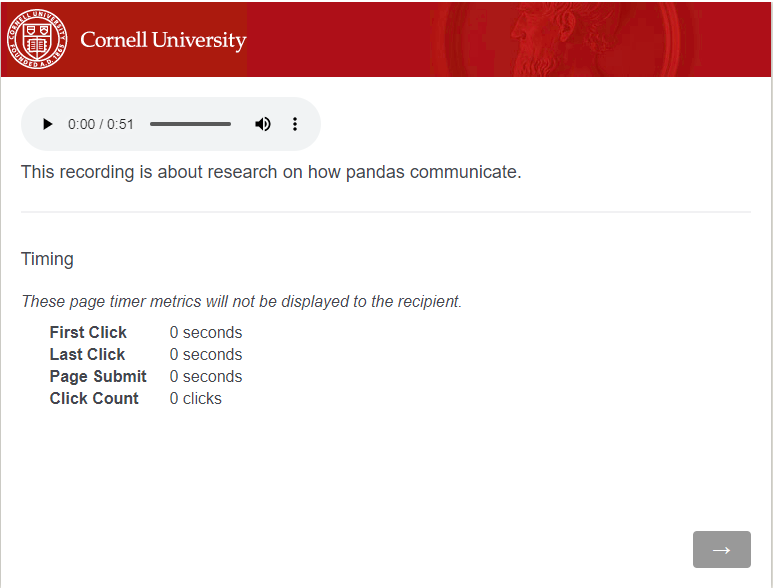}
  \end{figure}
%%%%%%%%%%%%%%%%%%%%%%%%%%%%%%%%%%%%%%%%%%%%%%%%%%%%%%%%%%%%%%%%%%%%%%%%%%%%% 
\vspace{0.3 cm}

\subsection{Treatment messages} \label{sec:Appendix_instructions}

\vspace{0.8 cm}
%%%%%%%%%%%%%%%%%%%%%%%%%%%%%%%%%%%%%%%%%%%%%%%%%%%%%%%%%%%%%%%%%%%%%%%%%%%%%
\FloatBarrier
\begin{figure}[H]
   \centering
    \caption{Treatment instructions text}
    \label{fig:instructions}
   \includegraphics[height=17cm,angle=0]{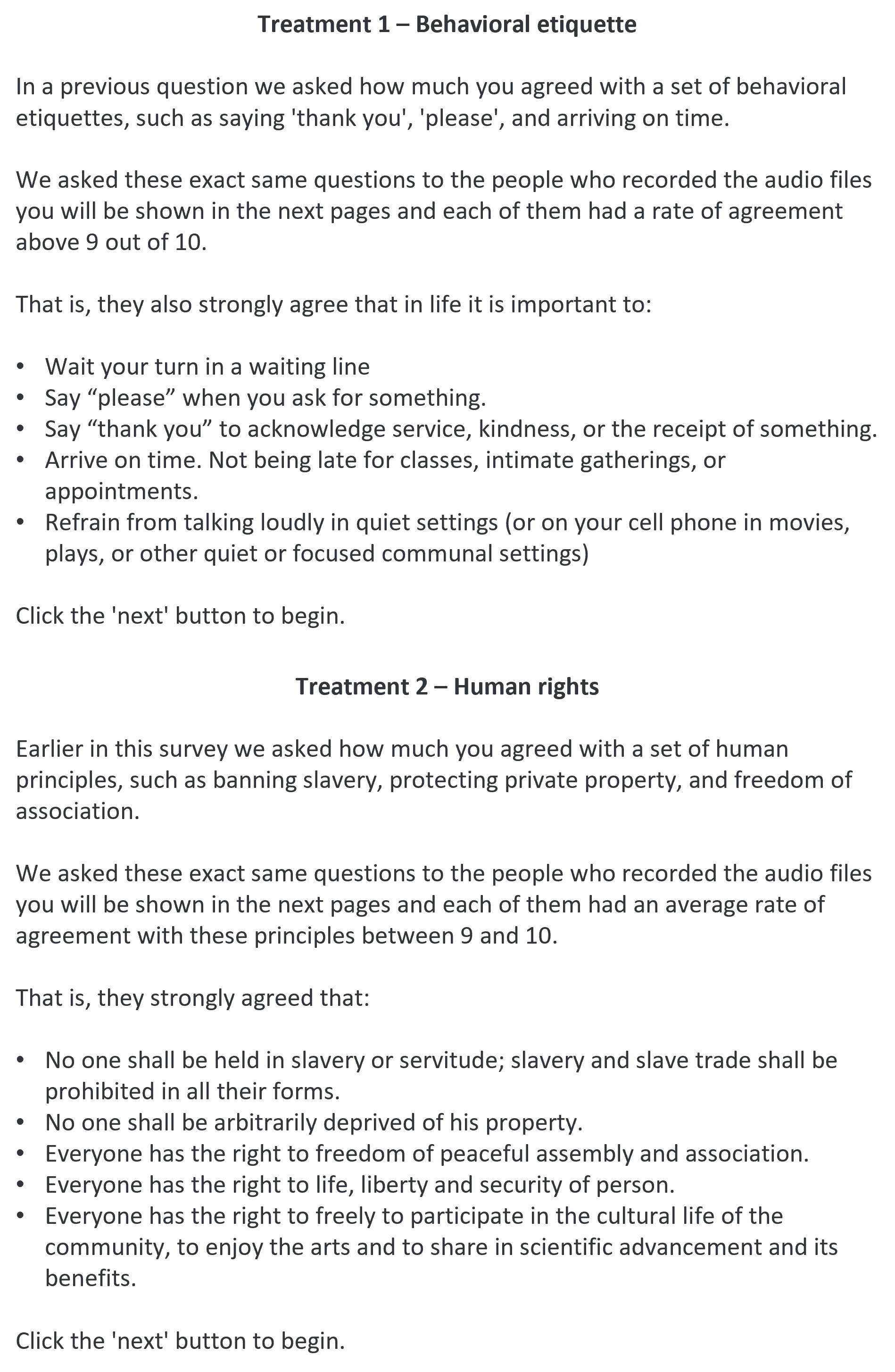}
  \end{figure}
%%%%%%%%%%%%%%%%%%%%%%%%%%%%%%%%%%%%%%%%%%%%%%%%%%%%%%%%%%%%%%%%%%%%%%%%%%%%% 
\vspace{0.3 cm}

\newpage

\subsection{Surveys} 

See Online Appendix \href{https://drive.google.com/file/d/1pzEoy8UYKtneE2wzvosEqMa7Lwrlg_DD/view?usp=sharing}{Link} for the full surveys conducted for (1) eliciting views on Human Rights and Basic Etiquette rules, (2) collecting recordings on views on Immigration, Abortion and Gun laws, (3) the main experiment.

\end{document}